\documentclass[trackchanges]{aastex7}

\begin{document}

\title{Stellar Rotation in the First Six Million Years: Rotational Velocities and Radii Estimates of T Tauri Stars in IC~5070 and IC~348}


\author[orcid=0000-0001-6389-5639,gname=Laurin, sname=Gray]{Laurin M. Gray}
\affiliation{Department of Astronomy, Indiana University, 727 East Third Street, Bloomington, IN 47405, USA}
\email[show]{grayla@iu.edu}  

\author[orcid=0000-0001-8283-4591,gname=Katherine, sname=Rhode]{Katherine L. Rhode} 
\affiliation{Department of Astronomy, Indiana University, 727 East Third Street, Bloomington, IN 47405, USA}
\email{krhode@iu.edu}

\author[orcid=0000-0001-6381-515X,gname=Luisa, sname=Rebull]{Luisa M. Rebull}
\affiliation{Infrared Science Archive (IRSA), IPAC, 1200 E. California Blvd., California Institute of Technology, Pasadena, CA 91125, USA}
\email{rebull@ipac.caltech.edu}

\begin{abstract}

We have acquired high-resolution optical spectroscopy for a sample of T Tauri stars (TTSs) in open clusters using Hydra on the WIYN 3.5m telescope, and present projected rotational velocities ($v$~sin~$i$ values) for 54 stars in IC~5070 and 99 stars in IC~348. We combine these with published values for stellar temperature, luminosity, rotation period, circumstellar disk status, and binarity; we are predominantly interested in how the last two factors may affect the rotation speeds of the stars. We find evidence to support theories that interaction with circumstellar disks may slow the rotation of TTSs compared to Class~III stars in both clusters. We also identify a higher fraction of slow-rotating Class~III stars in IC~348 compared to other clusters; we suggest that some fraction of these may be stars that recently lost their disks.  We find that a higher fraction of binary stars are rapid rotators compared to single stars, though not to a statistically significant degree. We also combine our $v$~sin~$i$ measurements with rotation periods to estimate projected stellar radii, which we compare to predictions from stellar evolution models using a maximum likelihood method. 
We continue to show that models with increasing starspot coverage reduce radius inflation and align better with published age estimates than models without starspots.
\end{abstract}


\section{Introduction} 



T Tauri stars (TTSs) are pre-main-sequence (PMS) stars with masses $\lesssim$2$-$3 $M_{\odot}$, and very young ages (typically $\lesssim$$10^7$ yr) (e.g., \citealt{Bertout1989, Herbst1994}).  They are extremely active and rapidly evolving stars, with strong magnetic fields that produce cooler, darker starspots on the surface, bipolar outflows and jets expelling material, and hot spots due to accretion from the circumstellar disk (e.g., \citealt{Mundt1983, Lada1985, Konigl1991, Shu1994a, Cody2014}).  TTSs are often separated into two classes primarily based on emission signatures resulting from the accretion of material onto the star from a circumstellar disk. Classical T Tauri Stars (CTTSs) often show excess emission in optical, UV, and/or X-ray wavebands, while weak-lined T Tauri Stars (WTTSs) have no emission lines like Ca II which might indicate accretion (e.g., \citealt{Walter1987, Bertout1989, Herbst1994, Hillenbrand1998b}).  \cite{Briceno2019} have also proposed a CWTTS classification, representing a transition stage between CTTS and WTTS where a weak or truncated disk may be present.  At all stages, TTSs show periodic variability, likely due to starspots on the rotating stellar surface, though these signatures can be masked in CTTSs by accretion-based flaring and occluding disks (e.g., \citealt{Herbst1987, Herbst1994, Cody2014}).  There is a parallel classification for protostars from \cite{Lada1987} which is based on observing the excess IR emission from the disk itself, where Class~II is for stars that show the disk signature and Class~III is for non-disked stars.  While there is significant overlap between Class~II stars and CTTSs, not all Class~II stars are necessarily accreting from their disks.  Likewise, some WTTSs may still have disks from which they accrete slowly or not at all.

Some TTSs, especially those with disk signatures, have been observed with slower rotation periods than expected, indicating that they have undergone significant angular momentum loss (e.g., \citealt{Vogel1981, Hartmann1986, Edwards1993, Clarke2000, Herbst2001, Herbst2002, CB2007}).  One proposed mechanism for angular momentum loss early in the stellar lifetime is disk-locking, where the stellar surface becomes synchronized to the slower rotation of the inner disk through magnetic field lines, along which the star is accreting disk matter (e.g., \citealt{Konigl1991, Shu1994a, Ostriker1995, Matt2005b, Bouvier2014}).  \cite{Serna2024} demonstrated that a combination of accretion-powered stellar winds and interaction between the magnetic fields lines may be responsible for the loss of angular momentum, regulating the rotation rate between the star and the disk.  Once the disk dissipates, the star will spin up; the estimated time scales for disk dispersal range from 1 to 5 Myr, or even as long as 10 Myr \citep{Edwards1993, Herbst2000, Rebull2004, Rebull2020, Fedele2010, Gallet2015, Hartmann2016, Pfalzner2024}.  Membership in a binary system is also known to affect individual stellar rotation rates over the lifetime of the member stars.  Stars in close binaries become tidally locked, so the angular momentum of the star is influenced by the kinematics of the binary system rather than magnetic braking and stellar winds \citep{Skumanich1972, Levato1974, Zahn1989}.  This has resulted in observations that stars in binary systems tend to have faster rotation rates than single stars, even at very young ages \citep{Stauffer2018, Kounkel2023}.  Accurate stellar ages are necessary to study the rotational evolution due to these effects, although they can be difficult to derive based on the rapid evolution of the stars.

Stellar evolution models are frequently used to estimate ages, but many observational studies have found evidence that PMS stellar radii are larger than predictions from models like \cite{Baraffe2015}.  The observed evidence includes direct radii measurements of individual low-mass PMS eclipsing binaries (e.g., \citealt{Lopez-Morales2007, Torres2010, Kraus2015, David2019, Smith2021}) as well as statistical studies of projected stellar radius for large samples of single stars (e.g., \citealt{Lanzafame2017, Jackson2018, Gray2024}).  ``Inflated radii" have been linked to increased magnetic activity, which would inhibit convection and increase starspot coverage; in this work, we explore the impacts of starspots on the radii of TTSs (e.g., \citealt{Gough1966, Feiden2013, Somers2015a, Somers2020, Kiman2024}).  Recent work by \cite{Gangi2022} and \cite{PerezPaolino2024} has suggested that large fractions of the surfaces of TTSs may be covered in starspots, often in excess of 50\%, which would have a significant effect on the the model predictions.  We can combine the projected rotation velocity with the rotation period to estimate the projected radius of a star, which is not dependent on the luminosity or temperature (e.g., \citealt{Rhode2001}).  This measurement is dependent on the inclination $i$ of the star's spin-axis with respect to the observer, which is unknown, and therefore represents a lower limit for the radii of individual stars; however, with a larger sample of measurements, we can statistically compare the average radii to those predicted by stellar evolution models \citep{Lanzafame2017, Jackson2018}.

In \cite{Gray2024}, we conducted a pilot study in the $\sim$3 Myr cluster NGC 2264, to develop our methods for studying rotational evolution in TTSs.  We now apply these methods to two clusters with ages $\lesssim$5 Myr, IC~5070 and IC~348, to study the rotational evolution of TTSs in the era before the majority of disks have dissipated or before stars that previously had disks have had time to spin up.  IC~5070, also known as the Pelican Nebula, is an emission nebula containing many YSO candidates.  It is often studied alongside the larger North America Nebula, as both are part of the HII region Westerhout 80 but appear separated by a foreground dust cloud known as L 985 \citep{Rebull2011, Froebrich2021}.  For this work, we chose to focus on IC~5070 because the stars with published rotation periods were concentrated in that region, 
allowing us to maximize the number of stars in a pointing.  Recent studies have estimated distances between $\sim$800 and 900 pc \citep{Bhardwaj2019, Kuhn2020, Fang2020}.  \cite{Kuhn2020} found that the YSOs in IC~5070 are almost all younger than 3 Myr, with most being closer to 1 Myr in age.  IC~348 is an open cluster with a well-studied population within the Perseus molecular cloud, only $\sim$315 pc away \citep{Herbig1998, Luhman2003, Pang2022}.  Previous estimates for IC~348's age were generally around 2$-$3 Myr old \citep{Luhman1998, Luhman2003, LEL2016, Wang2022}, but based on JWST observations of brown dwarfs in IC~348, \cite{Luhman2024} recently revised their estimate to 5 $\pm$ 2 Myr.  This is closer to the estimate of 6 Myr in \cite{Bell2013}, which was based on isochrone fitting to PMS stars.

We have collected high-resolution spectra for over 150 TTSs in IC~5070 and IC~348 with the WIYN 3.5m telescope, so that we can measure their $v$~sin~$i$.  We use this to explore factors that may impact the rotational evolution of the star such as disk interactions and binarity, and we also test radius predictions from stellar evolution models.  In Section \ref{sec:analysismethods}, we discuss our observations and the methods we use to measure rotation velocities, and in Section \ref{sec:sample}, we explain how we selected and classified our targets and present our final analysis samples.  In Section \ref{sec:vsini_distribs}, we discuss the $v$~sin~$i$ distributions for the stars based on their disk status and binarity.  In Section \ref{sec:rad_comp}, we present our maximum likelihood model and results for comparing measured radii to model predictions. In Section \ref{sec:conclusion}, we review our findings.  

\section{Observations and Measurements} \label{sec:analysismethods}

\subsection{Observations}
All spectroscopic observations were completed with the WIYN 3.5m telescope at Kitt Peak National Observatory, using the Hydra multifiber positioner and Bench spectrograph, which allows us to take high-resolution spectroscopic measurements across a 1$^\circ$ field of view.  Specific details about the target stars selected are located in Section \ref{sec:targets}.  Our instrument setup was identical to the one used in \cite{Gray2024}: we used the red Hydra fibers, which each have an aperture of 2$\arcsec$, and the 316@63.4 echelle grating, covering a spectral range of 6240$-$6540 {\AA} and centered on 6400 {\AA} at a spectral resolution of $R$ $\sim$ 21,500.  We took Thorium$-$Argon (ThAr) comparison lamp spectra, dome flats, zeros, and darks to accurately solve the wavelength solution and to correct for instrumental effects.  

Observations for IC~5070 were completed on 2024 May 11$-$12.  We observed two configurations, with around 50 objects and at least 14 sky fibers in each, for 3$-$3.5 hr each.  We collected spectra for 73 objects in IC~5070, with 25 objects observed in both configurations.    

Observations for IC~348 were taken in 2023 November, 2023 December, and 2024 November.  We observed four different configurations, each of which had $\sim$45$-$60 objects and at least 13 sky fibers.  The total exposure time for each configuration varies between 1.5 and 3.5 hr.  Spectra were observed for 129 objects, with 47 included in multiple configurations.  

\subsection{Data Reduction}
Data reduction followed the same procedure as described in \cite{Gray2024}, and is summarized here.  The spectroscopic data were reduced with IRAF \citep{Tody1986,Tody1993}.  Combined zero, dark, and dome flat images were created, and the dome flat, comparison lamp, and science images were zero subtracted, dark subtracted, and trimmed with CCDPROC.  We used DOHYDRA to extract apertures, apply the dome flat, calibrate the wavelength dispersion function using the ThAr spectra, and subtract the average sky spectrum from the object spectrum.  Flux due to emission from the surrounding nebula may be captured in the sky fibers, but the variable background is not expected to affect the rotation velocity measurements \citep{Rhode2001}.

We used the RVCORRECT IRAF task to perform a heliocentric Doppler motion correction on each exposure in a configuration to align with the first exposure; for the observations in this paper, all exposures within each configuration were taken on either the same night or two consecutive nights, so the difference in heliocentric Doppler motion across exposures is negligible.  Following this, we used SCOMBINE to scale, weight, and average together the exposures for each configuration into a single multispectrum image.  We did not combine the spectra of stars that were observed in multiple configurations.  We removed the ends of the spectra where the focus began to degrade, leaving us with a spectral range of 6275$-$6525 {\AA}.  We used CONTINUUM to continuum-normalize the spectra with a second-order spline3 function. 

\subsection{Measuring Radial and Rotational Velocities}

We used the IRAF task FXCOR to cross-correlate each object spectrum against a template spectrum of a narrow-lined star.  By performing a Fourier cross-correlation of a stellar spectrum against a template spectrum with a known radial velocity (RV), we can measure the RV from the peak location of the cross-correlation function (CCF) and projected rotational velocity ($v$~sin~$i$) from the width of the peak. 
While many factors such as turbulence can contribute to broadening of absorption lines, rotational broadening is the dominant source of line broadening for PMS stars over pressure broadening, especially at the resolution of our measurements \citep{Dahm2012}.  

Our template spectra are seven slow-rotating stars with spectral types ranging from G2V to M2V, which were previously observed with the WIYN 3.5m telescope and reduced as described in \cite{Gray2024}.  Each target star was cross-correlated against the template with the closest effective temperature.  To determine the relationship between the velocity of the star and the width of the CCF peak, we artificially broadened each template with a rotation profile to a range of values between 5 and 150 $\mathrm{km}\,\mathrm{s}^{-1}$.  Each broadened spectrum was cross-correlated against the original template and the resulting FWHM was measured.  We fit a fourth-order polynomial to the FWHM measurements and the corresponding rotation velocities of the spectra to create a calibration function.  

The signal-to-noise ratio of the cross-correlation peak is characterized by the Tonry$-$Davis parameter $R_{TD}$, and the uncertainty for the rotational velocity is inversely proportional to $(1 + R_{TD})$ \citep{Tonry1979}.  We used the relationship of $\pm$ $v$~sin~$i$$/(1 + R_{TD})$ defined in \cite{Hartmann1986} to estimate the uncertainty for individual measurements of $v$~sin~$i$.  

\begin{figure}
    \centering
    \includegraphics[width=0.4\linewidth]{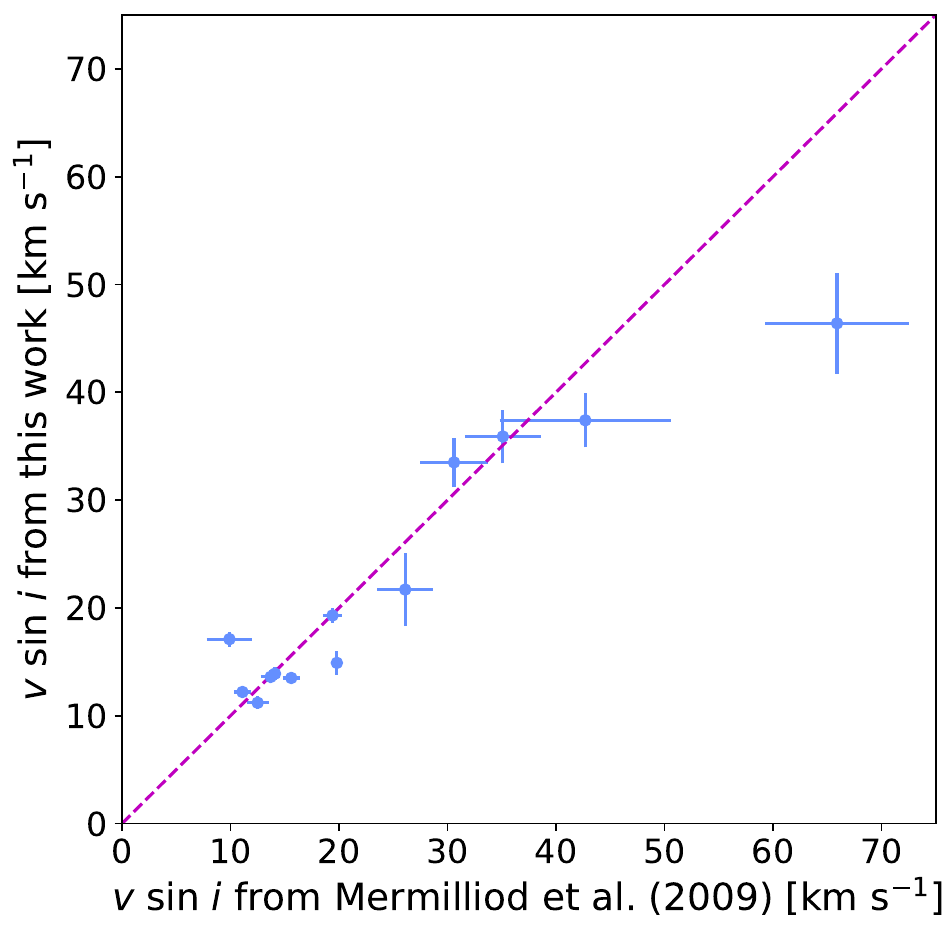}
    \caption{Plot of our measured $v$~sin~$i$ values vs. values for the same stars from \cite{Mermilliod2009}, to demonstrate the effectiveness of our measurement methods.  The magenta dashed line depicts a perfect 1-to-1 correlation.  For objects with $v$~sin~$i$ $\geq$ 11 $\mathrm{km}\,\mathrm{s}^{-1}$, we find strong agreement between the measurements, excluding vB~36 as discussed in the text.} 
    \label{fig:vsini_standards}
\end{figure}

Based on our explorations in \cite{Gray2024}, where we compared our measurements of $v$~sin~$i$ for a small sample of stars to those taken with higher resolution and precision in \cite{Mermilliod2009}, we determined that the absolute lower resolution limit for our measurements is around 11~$\mathrm{km}\,\mathrm{s}^{-1}$, although reported values slightly above this limit may still be upper limits rather than true values in less than ideal conditions.  We have added three more stars to this sample, for a total of 13 stars (see Table \ref{tab:vsini_standards}), and find that we still have strong agreement between our measurements and those of \cite{Mermilliod2009}.  A comparison of our $v$~sin~$i$ measurements to those from \cite{Mermilliod2009} is shown in Fig. \ref{fig:vsini_standards}.  While our $v$~sin~$i$ measurement for vB~36 significantly disagrees with the measurement in \cite{Mermilliod2009}, other literature values range from 40 to 51~$\mathrm{km}\,\mathrm{s}^{-1}$, which supports the validity of our measurement \citep{Glebocki2005, Cummings2017}.  When we perform a linear least-squares regression excluding vB~36 and He~347 (which is below our $v$~sin~$i$ resolution limit), we find a slope of 0.938 $\pm$ 0.083 with an intercept of 0.110 $\pm$ 1.991 and a Pearson correlation coefficient $r$ of 0.967.  For a strong agreement between measurements, we expect a slope in agreement with 1, an intercept in agreement with 0, and an $r$ $>$ 0.95, so we conclude that our calibration curves and methodology provide accurate measurements. 

\begin{deluxetable}{ccccccccc}
\tablecaption{Stars from \cite{Mermilliod2009} used for $v$~sin~$i$ measurement comparison}
\tablehead{\colhead{Star} & \colhead{RA} & \colhead{Dec} & \colhead{V\tablenotemark{a}} & \colhead{RV\tablenotemark{b}} & \colhead{RV\tablenotemark{c}} & \colhead{$v$~sin~$i$\tablenotemark{b}} & \colhead{$v$~sin~$i$\tablenotemark{c}} & \\ 
\colhead{} & \colhead{(deg)} & \colhead{(deg)} & \colhead{(mag)} & \colhead{($\mathrm{km}\,\mathrm{s}^{-1}$)} & \colhead{($\mathrm{km}\,\mathrm{s}^{-1}$)} & \colhead{($\mathrm{km}\,\mathrm{s}^{-1}$)} & \colhead{($\mathrm{km}\,\mathrm{s}^{-1}$)} } 

\startdata
He 347 & 49.40617 & 48.10798 & 10.55 & 8.67 $\pm$ 0.36 & 7.78 $\pm$ 1.11 & 9.9 $\pm$ 2.1 & 17.1 $\pm$ 0.7 \\
AK 511 & 129.30384 & 17.25395 & 8.94 & 42.75 $\pm$ 0.39 & 44.19 $\pm$ 0.93 & 11.1 $\pm$ 0.8 & 12.2 $\pm$ 0.5 \\
AK 1159 & 130.71805 & 17.1337 & 8.76 & 20.5 $\pm$ 0.31 & 19.24 $\pm$ 1.3 & 12.5 $\pm$ 1 & 11.2 $\pm$ 0.6 \\
VL 1353 & 130.65319 & 18.3888 & 10.08 & 35.99 $\pm$ 0.47 & 36.51 $\pm$ 1.2 & 13.7 $\pm$ 0.9 & 13.6 $\pm$ 0.6 \\
Malm 30.252 & 188.75103 & 30.19252 & 8.62 & -1.91 $\pm$ 0.2 & -2.9 $\pm$ 1.08 & 14.1 $\pm$ 0.5 & 13.9 $\pm$ 0.6 \\
He 299 & 48.99553 & 50.4051 & 11.15 & -1.82 $\pm$ 0.48 & -2.78 $\pm$ 0.87 & 15.6 $\pm$ 0.8 & 13.5 $\pm$ 0.5 \\
He 334 & 49.24788 & 49.9265 & 10.37 & -2.47 $\pm$ 0.72 & -3.63 $\pm$ 1.17 & 19.4 $\pm$ 0.9 & 19.3 $\pm$ 0.7 \\
Tr 19 & 183.10367 & 27.38006 & 8.06 & 0.66 $\pm$ 0.24 & -1.85 $\pm$ 2.06 & 19.8 $\pm$ 0.4 & 14.9 $\pm$ 1.1 \\
Tr 101 & 185.92081 & 26.9799 & 8.36 & -0.17 $\pm$ 0.4 & -0.25 $\pm$ 5.14 & 26.1 $\pm$ 2.6 & 21.7 $\pm$ 3.4 \\
KW 411 & 130.40066 & 19.14259 & 9.32 & 35.6 $\pm$ 1.42 & 32.11 $\pm$ 3.06 & 30.6 $\pm$ 3.1 & 33.5 $\pm$ 2.3 \\
Tr 36 & 184.03484 & 25.76033 & 8.07 & 0.03 $\pm$ 0.79 & 0.31 $\pm$ 3.2 & 35.1 $\pm$ 3.5 & 35.9 $\pm$ 2.5 \\
Art I.247 & 126.41523 & 18.31896 & 10.22 & 36.06 $\pm$ 1.47 & 32.76 $\pm$ 3.23 & 42.7 $\pm$ 7.9 & 37.4 $\pm$ 2.5 \\
vB 36 & 65.38497 & 18.41744 & 6.79 & 40.73 $\pm$ 2.42 & 37.63 $\pm$ 5.83 & 65.9 $\pm$ 6.6 & 46.4 $\pm$ 4.7 \\
\enddata

\label{tab:vsini_standards}
\tablecomments{Right ascension and declination are given in degrees, using coordinates from Gaia DR3 \citep{GaiaEDR3-2021a}.}
\tablenotetext{a}{$V$-band magnitude is converted from Gaia DR3 $G$-band and $G_{BP}-G_{RP}$, using the conversions from \cite{Riello2021}.}
\tablenotetext{b}{\cite{Mermilliod2009}}
\tablenotetext{c}{This work.}
\end{deluxetable}

\section{Analysis Sample Selection} \label{sec:sample}

\subsection{Target Sample Selection} \label{sec:targets}
We relied on photometric and spectroscopic catalogs from the literature to compile a target sample of low-mass PMS stars in IC~5070 and IC~348.  These catalogs often contained relevant auxiliary information such as spectral types or temperatures, rotation periods, RVs, and disk classifications.  For IC~5070, we began by cross-matching our two most robust membership catalogs, spectroscopically confirmed members from \cite{Fang2020} and astrometrically confirmed members from \cite{Kuhn2020}, and then added photometry, rotation periods, and disk classifications from \cite{Rebull2011}, \cite{Bhardwaj2019}, \cite{Froebrich2021}, and \cite{Hillenbrand2022}.  For IC~348, we used \cite{LEL2016} as the base for our catalog, 
then added \cite{HMW2000}, \cite{Luhman2003}, \cite{CHW2004}, \cite{LNB2005}, \cite{KKB2005}, \cite{CB2006}, \cite{Lada2006}, \cite{Nordhagen2006}, and \cite{Fritz2016} for rotation periods and disk classifications, and RV and $v$~sin~$i$ measurements from APOGEE from \cite{Cottaar2014, Cottaar2015} and \cite{Kounkel2019}.  For each catalog, we first cross-matched the contents to the Gaia DR3 archive within 2$\arcsec$, noting any additional objects within 2$\arcsec$ but selecting the closest coordinate match \citep{GaiaEDR3-2021a}.  This allowed us to have a common astrometric solution for all objects in our database.  We excluded any objects that did not have a Gaia object within 2$\arcsec$.  For each object, we also used the relations in \cite{Riello2021} to calculate a $V$-band magnitude from the Gaia DR3 photometry.  When multiple Gaia objects were within 2$\arcsec$ of the original coordinate, we checked whether they were classified as a multi-object system in Simbad \citep{Simbad2000}; if they were, we noted the corresponding component and made sure the correct Gaia number was assigned.  

To select stars from the same range of masses between clusters, we applied magnitude and effective temperature cuts as appropriate.  Our maximum effective temperature was defined as the temperature of a 1.4 solar mass star at the upper age limit of the cluster (3 Myr for IC~5070 and 6 Myr for IC~348), as predicted by \cite{Baraffe2015}. For IC~5070, we selected objects with apparent $I_C$-band magnitudes between 13.5 and 16.5 or $V$-band magnitudes between 15 and 20.5, and effective temperatures below 4670~K. For IC~348, we accepted objects with apparent $I_C$-band magnitudes between 12 and 16.5 or $V$-band magnitudes from 13 to 20.5 and our effective temperature cutoff was 4920~K.


\subsection{Adopted Values from the Literature} \label{sec:adoption}

For stars with known effective temperatures ($T_{\rm eff}$), rotation periods ($P$), and bolometric luminosities ($L_*$), we can calculate the radius $R_*$ = $\sqrt{L_*/4 \pi \sigma_{SB} T_{eff}^4}$, and then the equatorial rotation velocity $V_{eq}$ = 2$\pi R_*/P$.  We also need a $T_{\rm eff}$ estimate to select an appropriate spectral template for cross-correlation.

To assign temperatures to stars as consistently as possible, we preferred to adopt a spectral type whenever available and use the relationships in \cite{PM2013} to derive an appropriate effective temperature.  \cite{PM2013} suggests that PMS stars may have slightly cooler surface temperatures than their MS counterparts of the same spectral type, particularly in the range of G5$-$K6.  For IC~5070, we adopted spectral types from \cite{Fang2020}, followed by $T_{\rm eff}$ from \cite{Bhardwaj2019}. Spectral types for IC~348 were primarily adopted from \cite{LEL2016}, with temperatures from \cite{Cottaar2014} and \cite{Kounkel2019} when the spectral type was unknown.

Rotation periods for IC~5070 were adopted from \citeauthor{Hillenbrand2022} (\citeyear{Hillenbrand2022}; only adopting those classified as periodic or quasi-periodic), followed by those from \cite{Froebrich2021} and \cite{Bhardwaj2019}.  For IC~348, one of us (L.M.R.) obtained rotation periods from the Transiting Exoplanet Survey Satellite (TESS; \citealt{Ricker2014}) and the Zwicky Transient Facility (ZTF; \citealt{Kulkarni2018}) using the methods in \cite{Rebull2018}.  These frequently confirmed previously published periods from \cite{Fritz2016}, \cite{CB2006}, \cite{Nordhagen2006}, \cite{KKB2005}, \cite{LNB2005}, and/or \cite{CHW2004}; when we were unable to identify a period from the TESS or ZTF data, we adopted one from the literature in the order listed.

For IC~5070, we adopted the model luminosities from \cite{Fang2020}, which were derived from spectral energy distribution (SED) fitting to a veiled and reddened model atmosphere of the same spectral type.  For IC~348, bolometric luminosities have been published in \cite{Luhman2003} for about half of our target sample.  These luminosities were derived using colors and bolometric corrections from \cite{Kenyon1995} for main-sequence (MS) dwarfs.  However, since we are using the cooler temperatures derived for PMS stars from \cite{PM2013} and also wanted to maximize the number of stars for which we can calculate $V_{eq}$, for the sake of consistency, we estimated $L_{bol}$ for all target stars based on the method described in \cite{Rebull2018}.  We begin with a spectral type or effective temperature and a $V$-band magnitude from Gaia DR3 for all of our stars, along with archival photometry for $JHK_s$ for the majority from \cite{LEL2016}.  First, we calculated $E(J - K_s)$ using the $(J - K_s)$ predicted by \cite{PM2013} for the star's spectral type and used the reddening law from \cite{Indebetouw2005}, $A_{K_s}$ = 0.667$E(J - K_s)$.  Then, we used the relations from \cite{Cardelli1989}, $A_{K_s}$ = 0.114$A_V$, to calculate $E(V - K_s)$ and $(V-K_s)_0$.  This allowed us to avoid the ``quantization" that results from using the expected $(V - K_s)$ color directly for 43\% of our sample.  However, if the dereddening from the $(J - K_s)$ color was unphysical (the color excess was $<$0) or the photometry was not available, we calculated $E(V - K_s)$ using the $(V - K_s)$ colors from \citeauthor{PM2013} (\citeyear{PM2013}; 52\% of the sample).  The mode $E(V - K_s)$ derived was 1.86.  For the remaining 5\% of stars that we could not estimate $E(V - K_s)$ for, we applied this mode $E(V - K_s)$ and corresponding $A_V$.  We calculated $V_0$ from the $V$-band magnitude derived from Gaia DR3, applied bolometric corrections according to spectral type from \cite{PM2013}, and used a distance of 315 pc to estimate the bolometric luminosity.  For the 113 objects that also had bolometric luminosities reported in \cite{Luhman2003}, our estimates were slightly lower by an average of 0.11 $L_{\odot}$.  

As discussed in \citet{Gray2024}, classifying TTSs can be a complicated process.  The main diagnostic methods focus on either evidence for disk presence through IR excess emission or accretion signatures (usually UV excess emission or $H\alpha$ equivalent widths, hereafter EWs), which are related but not always co-occurring \citep{Venuti2018}.  For example, a TTS may show evidence of a disk through IR excess, but may be going through a period of low accretion and would not be identified as disk-bearing from an $H\alpha$ EW measurement alone.  Many studies only use one of these diagnostic methods and so do not distinguish between a disked star with accretion signatures and a disked star without, for example.  Additionally, accretion rates and IR excess emission are variable for TTSs, so the classification may be different for measurements taken at different times \citep{Venuti2018}.  \cite{Rice2012} found that, for 25 YSOs with a detected near-IR (NIR) excess, 7 of them only showed an NIR excess between 15\% and 65\% of the time, and 3 of those did not have an NIR excess in the time-averaged $JHK$ data.  Finally, \cite{Briceno2019} have proposed an additional CWTTS classification based on H$\alpha$ EWs, which may indicate stars in the process of transitioning to lower accretion activity or a weaker/truncated disk.  To classify stars in our sample as Class~II, Class~III, CTTS, and CWTTS, we relied on literature classifications derived from a mix of IR excess measurements and $H\alpha$ EWs. We do not specifically classify WTTSs; these stars would be a mix of Class~II and Class~III, but our primary interest is in comparing stars with disks (Class~II) to those without disks (Class~III), and stars with accretion (CTTS) against those that are unlikely to have accreted material recently (Class~III).  For IC~5070, we did not have any $H\alpha$ EW measurements from the literature, so our classifications were based on those from \cite{Rebull2011}, \cite{Bhardwaj2019}, and \cite{Fang2020}, all of which relied on measurements of IR excess and so stars were only classified as Class~II or Class~III.  For IC~348, we first identified any possible transition CWTTSs by an $H\alpha$ EW measurement from \cite{Herbig1998} or \cite{Luhman2003} according to the system from \cite{Briceno2019}.  Then, the remainder were classified as Class~II or Class~III based on IR excess measurements from \cite{Lada2006}, \cite{LEL2016}, \cite{Kounkel2019}, and L. M. Rebull (2025, private communication).  Sources with high H$\alpha$ emission according to \cite{Briceno2019} were additionally classified as CTTS; with one exception, these were all also Class~II objects.  We are limited in the number of stars that have H$\alpha$ EW measurements in the literature, but \cite{Venuti2018} found that less than 5\% of the over 300 stars without IR excesses in their sample had accretion signatures.  Of the 21 single and possible binary stars with no IR excess emission in our own sample, the one that showed H$\alpha$ emission (star ID 89) was removed from the Class~III sample and reclassified as CTTS, but not Class~II. 

\subsection{Membership and Classification} \label{sec:classifying}

In order to properly evaluate the rotational properties of a cluster, we must ensure that we use good quality measurements, and take care to identify binary stars and remove stars that are not cluster members.  We accomplish this by relying on a combination of qualitative and quantitative evaluation of the CCF peaks, auxiliary data, and comparisons of measurements from different times when they are available.

We first removed measurements with $R_{TD}$~$<$~2.5, to remove CCF peaks that were potentially random noise rather than a true measurement. Next, we started identifying binary stars using the shape of the CCF peak.  Stars with asymmetries (e.g., side lobes or slightly non-Gaussian shapes) in the CCF peak were labeled as ``possible binaries" while stars with clear multiple Gaussian peaks were marked as ``likely binaries."  After this initial pass through our data, we began incorporating information from the literature.  Stars that had been identified as binaries in the literature (\citealt{Bhardwaj2019}, \citealt{Froebrich2021} for IC~5070; \citealt{Nordhagen2006}, \citealt{Kounkel2019} for IC~348) were reclassified as ``likely binaries" if necessary. 

The renormalized unit weight error (RUWE) from Gaia, which quantifies how well the astrometric observations match the model for a single star, is often used to identify close binaries that are not resolved by Gaia \citep{Ziegler2018}.  Stars with RUWE near 1 are considered well fit, while stars with RUWE greater than some threshold (typically 1.4) are more likely to be part of a binary system. However, \cite{Fitton2022} suggests that the traditional cutoff of 1.4 that was derived from field stars may not be appropriate for PMS stars, particularly those that have protoplanetary disks. \cite{Fitton2022} found that 95\% of single disk-bearing stars had RUWE $\leq$ 2.5 and that the 95th percentile for single disk-free PMS stars was 1.6.  We classified stars with RUWE exceeding these thresholds according to their adopted disk classifications as ``likely binaries."  

We also looked at whether the rotation periods reported were similar or in disagreement, or if multiple periods had been noted; if they were, we marked those stars as ``possible binaries."  Single G or K stars may show multiple periods as a result of differential rotation and/or evolving starspot groups, but multiple periods are more likely to indicate a binary for M stars \citep{Rebull2016b, Rebull2017, Stauffer2016}.  For a single star, we expect the RV and $v$~sin~$i$ to be fairly consistent whenever they are measured, so significant differences may indicate that a binary system has been observed at a different point in its orbital cycle.  We calculated a weighted average for measurements from the same observing run, as we saw that measurements taken only $\sim$1 day apart were consistent.  For IC~5070, we only had one observing run and no RV or $v$~sin~$i$ measurements from the literature, which made it more difficult to identify binaries in that cluster beyond what was described previously.  For IC~348, we were able to compare the averaged measurements for each observing run to each other and to the literature \citep{Nordhagen2006, Cottaar2014, Cottaar2015, Kounkel2019}.  Stars with more than 2$\sigma$ difference in their RVs, or with more than 1$\sigma$ disagreement and asymmetry in their CCFs were marked ``likely binaries," and we noted whether the $v$~sin~$i$ measurements were above or below the resolution limit.  We calculated a weighted average for RV and $v$~sin~$i$ with all of our measurements for the remaining ``single" and ``possible binary" stars.

Finally, we created the samples we would use for analysis.  For stars where $v$~sin~$i$ or $V_{eq}$ $<$ 11 $\mathrm{km}\,\mathrm{s}^{-1}$, we marked them as having $v$~sin~$i$ with an upper limit of 11 $\mathrm{km}\,\mathrm{s}^{-1}$.  For stars where $v$~sin~$i$ $>$ $V_{eq}$, if we were confident in the $v$~sin~$i$ measurement (ex. it was consistent or had high $R_{TD}$), we evaluated other possible periods to see if they may be more appropriate and recalculated $V_{eq}$.  Otherwise, if it was already a ``possible binary" from the CCF, we reclassified it as a ``likely binary," with suspicion that the $v$~sin~$i$ may be overestimated from line-blending.  While our target sample was built out of stars that had been identified as cluster members primarily through their astrometry, proper motions, and photometry, we can now use RV to confirm cluster membership.  Following \cite{Furesz2006}, we define membership as having an RV within 4$\sigma$ of the mean RV. For IC~5070, we calculated the average RV using only objects identified as single stars: -16.34 $\mathrm{km}\,\mathrm{s}^{-1}$ with a velocity dispersion $\sigma$ = 2.55 $\mathrm{km}\,\mathrm{s}^{-1}$. Three objects had RV more than 4$\sigma$ away from the mean (RV $\sim$-32$-$-36 $\mathrm{km}\,\mathrm{s}^{-1}$), and all had been identified as ``likely binaries" through other methods.  For IC~348, \cite{Cottaar2015} measured an average RV of 15.37 $\mathrm{km}\,\mathrm{s}^{-1}$ with $\sigma$ = 0.72 $\mathrm{km}\,\mathrm{s}^{-1}$.  When we use this to establish the bounds of membership, we find that our single objects have an average RV of 16.04 $\mathrm{km}\,\mathrm{s}^{-1}$ with $\sigma$ = 1.28 $\mathrm{km}\,\mathrm{s}^{-1}$, which is in good agreement with the \cite{Cottaar2015} estimate.  For objects identified as ``likely binaries" to be considered a member of the cluster, they must have at least one RV measurement within the membership range, or, if multiple CCF peaks or measurements are present, the RV measurements must span the range.  

\subsection{Samples and Subsamples for Analysis}

Our final samples of low-mass PMS stars consist of 54 stars in IC~5070 and 99 stars in IC~348, listed in Tables \ref{tab:NAP_targets} and \ref{tab:IC348_targets}, respectively.  We have classified our stars to identify single stars, possible binaries, and likely binaries, according to the process described in Section \ref{sec:classifying}.  IC~5070 has 22 single stars, 13 possible binaries, and 19 likely binaries.  IC~348 has 29 single stars, 31 possible binaries, and 39 likely binaries.  We also create a Resolution-Limited Sample of single stars and possible binaries with $v$~sin~$i$ and $V_{eq}$ (if known) $\geq$ 11 $\mathrm{km}\,\mathrm{s}^{-1}$.

The sample size for IC~5070 is small, making statistical analyses difficult, so we have also used the literature to create a comparable sample of stars from the Orion Nebula Cluster (ONC), which is considered to be about the same age as IC~5070 (1$-$3 Myr) and will therefore be in a similar stage of evolution \citep{Megeath2016, DaRio2017, Kroupa2018}. In Section \ref{sec:rho_results}, we provide further justification for combining these samples based on age. We used $v$~sin~$i$, disk classifications, and rotation periods from \cite{Rhode2001}, and augmented it with rotation periods and disk classifications from \cite{Serna2021}.  While \cite{Serna2021} does report $v$~sin~$i$ values from \cite{Kounkel2019}, the \cite{Rhode2001} observations were taken with the same instrument and nearly identical instrumental setup as our own, so for consistency, we adopt the $v$~sin~$i$ values reported by \cite{Rhode2001}.  However, we do note significant differences between $v$~sin~$i$ values reported in both papers to identify likely binary stars, as well as employing the other binary identification methods mentioned in Section \ref{sec:classifying}, such as RUWE from Gaia DR3 and notes on the CCF and binarity from \cite{Rhode2001} and \cite{Serna2021}.  The temperatures used by \cite{Rhode2001} originated from spectral types in \cite{Hillenbrand1997} converted to temperatures with the scale of \cite{Cohen1979}.  To be consistent with the analysis for our other clusters, we reconverted the spectral types to effective temperatures using the \cite{PM2013} scale.  For the disk classifications, we followed a similar procedure described for IC~348 in Section \ref{sec:adoption} with the $H\alpha$ EWs from \cite{Serna2021} and NIR excess and Ca II emission-line strength from \cite{Rhode2001} to classify stars as CWTTS, Class~II, Class~III, and CTTS.  There were an additional nine stars that were not classified as Class~II or Class~III according to the classification scheme from \cite{Rhode2001}, but which had H$\alpha$ EWs that are high enough to be CTTS.   
This results in an additional 111 single stars, 25 possible binaries, and 53 likely binaries from the ONC.  Table \ref{tab:subsamples} contains a summary of the various samples used in the following analysis.

\begin{deluxetable}{ccccccccccccc}[h]
\tablewidth{1.\columnwidth}
\tabletypesize{\footnotesize}
\tablecaption{Measured Values of RV and $v$~sin~$i$ for Stars in IC~5070, with Auxiliary Data Adopted from the Literature}
\label{tab:NAP_targets}
\tablehead{\colhead{ID} & \colhead{Gaia DR3 ID} & \colhead{R.A.} & \colhead{Decl.} & \colhead{$T_{\rm eff}$\tablenotemark{a}} & \colhead{Per\tablenotemark{a}} & \colhead{Refs} & \colhead{$L_{\rm bol}$\tablenotemark{a}} & \colhead{$V_{eq}$\tablenotemark{b}} & \colhead{Disk\tablenotemark{a}} & \colhead{Binarity\tablenotemark{c}} & \colhead{RV} & \colhead{$v$~sin~$i$} \\ 
\colhead{} & \colhead{} & \colhead{(deg)} & \colhead{(deg)} & \colhead{(K)} & \colhead{(days)} & \colhead{} & \colhead{($L_{\odot}$)} & \colhead{($\mathrm{km}\,\mathrm{s}^{-1}$)} & \colhead{} & \colhead{} & \colhead{($\mathrm{km}\,\mathrm{s}^{-1}$)} & \colhead{($\mathrm{km}\,\mathrm{s}^{-1}$)} } 

\startdata
38  & G2067063447700277504 & 312.52521 & 44.29692 & 3160 & 3.952 & 1,2    & 0.467 & 29.19  & III      & A      & -20.12 $\pm$ 3.56 & $<$11  \\
50  & G2163139941968363520 & 312.65575 & 44.34806 & 3602 & 6.048 & 1,2    & 2.562 & 34.38  & II      & A      & -13.61 $\pm$ 5.81 & 11.5 $\pm$ 2.2 \\
52  & G2163140934100563968 & 312.65814 & 44.42010  & 3835 & 1.422 & 1,3    & 4.891 & 178.24 & II      & b      & -22.85 $\pm$ 6.58 & $<$11 \\
55  & G2163139873248885632 & 312.66853 & 44.33934 & 3160 & 5.035 & 1,2    & 0.298 & 18.30   & II      & B      & -18.20  $\pm$ 6.22 & $<$11 \\
72  & G2163139804529410944 & 312.70289 & 44.34812 & 3780 & 10.79 & 1,2    & 0.971 & 10.77  & II      & A      & -19.60  $\pm$ 1.57 & $<$11 \\
81  & G2163139770169674624 & 312.72395 & 44.35510  & 3372 &       & 1      & 1.784 &        & II      & B      & -18.08 $\pm$ 3.06 & $<$11 \\
83  & G2163156056685634944 & 312.72580  & 44.63561 & 3928 & 9.547 & 1,2    & 1.025 & 11.58  & II      & A      & -17.55 $\pm$ 1.17 & $<$11 \\
100 & G2067058740416252416 & 312.74312 & 44.24230  & 4373 & 4.878 & 1,2    & 1.868 & 24.70   & II      & A      & -13.35 $\pm$ 1.13 & 18.5 $\pm$ 0.6 \\
101 & G2067058740416252544 & 312.74341 & 44.24562 & 3686 & 3.422 & 1,2    & 3.137 & 64.21  &        & B      & -13.77 $\pm$ 3.00 & 14.8 $\pm$ 1.3 \\
102 & G2163136402915307136 & 312.74460  & 44.29188 & 3916 & 7.22  & 1,2    & 1.256 & 17.06  & II      & B      & -17.38 $\pm$ 3.13 & 17.1 $\pm$ 1.6 \\
\enddata

\tablecomments{R.A. and decl. are given in degrees, using coordinates from Gaia DR3 \citep{GaiaEDR3-2021a}.  A sample of the table is shown here; the full version is available online.}
\tablenotetext{a}{Compiled from the literature.  See Section \ref{sec:adoption} for a detailed discussion of how $T_{\rm eff}$, period, $L_{\rm bol}$, and disk classification were selected.  Temperature and rotation period adopted from (1) \cite{Fang2020}, (2) \cite{Hillenbrand2022}, (3) \cite{Bhardwaj2019}, and (4) \cite{Froebrich2021}.}
\tablenotetext{b}{Calculated from the adopted temperature, period, and luminosity.}
\tablenotetext{c}{``A" indicates a star that has been classified as a single star system, ``b" indicates a ``possible" binary system, and ``B" indicates a ``likely" binary system.  For a full discussion on classifications, see Section \ref{sec:classifying}.}

\end{deluxetable}

\begin{deluxetable}{ccccccccccccc}[h]
\tablewidth{1.\columnwidth}
\tabletypesize{\footnotesize}
\tablecaption{Measured Values of RV and $v$~sin~$i$ for Stars in IC~348, with Auxiliary Data Adopted from the Literature}
\label{tab:IC348_targets}
\tablehead{\colhead{ID} & \colhead{Gaia DR3 ID} & \colhead{R.A.} & \colhead{Decl.} & \colhead{$T_{\rm eff}$\tablenotemark{a}} & \colhead{Per\tablenotemark{a}} & \colhead{Refs} & \colhead{$L_{\rm bol}$\tablenotemark{a}} & \colhead{$V_{eq}$\tablenotemark{b}} & \colhead{Disk\tablenotemark{a}} & \colhead{Binarity\tablenotemark{c}} & \colhead{RV} & \colhead{$v$~sin~$i$} \\ 
\colhead{} & \colhead{} & \colhead{(deg)} & \colhead{(deg)} & \colhead{(K)} & \colhead{(days)} & \colhead{} & \colhead{($L_{\odot}$)} & \colhead{($\mathrm{km}\,\mathrm{s}^{-1}$)} & \colhead{} & \colhead{} & \colhead{($\mathrm{km}\,\mathrm{s}^{-1}$)} & \colhead{($\mathrm{km}\,\mathrm{s}^{-1}$)} } 

\startdata
10 & G216676981211381632 & 56.18631 & 32.06734 & 3700 & 3.045  & 1,4 & 1.811 & 54.41 & C/II & B & 18.89 $\pm$ 1.71 & $<$11  \\
17 & G216678424320382848 & 56.16131 & 32.14499 & 4550 & 2.41   & 1,4 & 3.124 & 59.71 & III & B & 52.67 $\pm$ 3.17 & 29 $\pm$ 2   \\
20 & G216678389960647552 & 56.13140 & 32.14580 & 4760 & 2.24   & 1,4 & 1.687 & 43.14 & III & b & 17.47 $\pm$ 3.75 & 44.8 $\pm$ 2.7 \\
23 & G216678321241170432 & 56.15784 & 32.13447 & 3970 & 7.904  & 1,4 & 1.445 & 16.27 & C/II & A & 15.90 $\pm$ 3.27 & 15 $\pm$ 1.5 \\
25 & G216678115082741248 & 56.16351 & 32.12652 & 4550 & 5.203  & 1,8 & 0.779 & 13.82 & III & A & 14.56 $\pm$ 1.62 & $<$11  \\
26 & G216678115082741632 & 56.16023 & 32.12656 & 4020 & 5.218  & 1,4 & 1.245 & 22.30 & II & A & 16.99 $\pm$ 3.38 & 18.8 $\pm$ 1.8 \\
27 & G216676775052953216 & 56.15825 & 32.05822 & 4020 & 8.711  & 1,4 & 1.292 & 13.61 & C/II & B &                  &     \\
29 & G216702647935913344 & 56.25729 & 32.24099 & 4330 & 16.824 & 1,4 & 1.006 & 5.36  & III & A & 15.88 $\pm$ 1.54 & $<$11  \\
30 & G216681516696833408 & 56.12387 & 32.17771 & 3940 & 4.6    & 1,5 & 1.006 & 23.68 & C/II & A & 14.30 $\pm$ 1.30 & 14.1 $\pm$ 0.6 \\
31 & G216681654135789696 & 56.09007 & 32.17711 & 3970 & 7.56   & 1,5 & 0.710 & 11.92 & C/II & A & 15.04 $\pm$ 3.54 & $<$11  \\
\enddata

\tablecomments{R.A. and decl. are given in degrees, using coordinates from Gaia DR3 \citep{GaiaEDR3-2021a}.  A sample of the table is shown here; the full version is available online.}
\tablenotetext{a}{Compiled from the literature.  See Section \ref{sec:adoption} for a detailed discussion of how $T_{\rm eff}$, period, $L_{\rm bol}$, and disk classification were selected.  Temperature and rotation period adopted from (1) \cite{LEL2016}, (2) \cite{Cottaar2014}, (3) \cite{Kounkel2019}, (4) this work, (5) \cite{Fritz2016}, (6) \cite{CB2006}, (7) \cite{Nordhagen2006}, (8) \cite{KKB2005}, (9) \cite{LNB2005}, and (10) \cite{CHW2004}.}
\tablenotetext{b}{Calculated from the adopted temperature, period, and luminosity.}
\tablenotetext{c}{``A" indicates a star that has been classified as a single star system, ``b" indicates a ``possible" binary system, and ``B" indicates a ``likely" binary system.  For a full discussion on classifications, see Section \ref{sec:classifying}.}
\end{deluxetable}

\begin{deluxetable}{lllllll}[h]
\tablewidth{0.5\columnwidth}
\tabletypesize{\footnotesize}
\tablecaption{Summary of Analysis Samples}
\label{tab:subsamples}
\tablehead{\colhead{Sample} & \colhead{Sub-sample} & \colhead{Conditions} & \colhead{Used in:} & \colhead{IC~5070} & \colhead{IC~5070/ONC} & \colhead{IC~348} } 

\startdata
Total & & RV member & & 54 (30) & 243 (105) & 99 (33) \\
Resolution-Limited Sample & & Singles \& possible binaries, & Section \ref{sec:inc} & 15 & 66 & 14 \\
 & &  $v$~sin~$i$ \& $V_{eq}$ (if known) $\geq$ 11 & & & & \\
Class~II & & Singles \& possible binaries, IR excess & Section \ref{sec:disks} & 27 (13) & 94 (39) & 17 (7)  \\
Class~III & & Singles \& possible binaries, no IR excess & Section \ref{sec:disks} & 8 (2) & 28 (18) & 40 (6)  \\
CTTS & & Singles \& possible binaries,  & Section \ref{sec:disks} & N/A & 25 (12) & 8 (3) \\
 & & H$\alpha$ EW according to \cite{Briceno2019} & & & & \\
CWTTS &  & Singles \& possible binaries, & Section \ref{sec:disks} & N/A & 6 (2) & 2 (1)   \\
 & & H$\alpha$ EW according to \cite{Briceno2019} & & & & \\
Quality singles &  & Singles, $R_{TD}$ $\geq$ 4.4 or 7, & Section \ref{sec:binaries} & 8 & 31 & 5  \\
 & &  $v$~sin~$i$ \& $V_{eq}$ (if known) $\geq$ 11 & & & & \\
Quality likely binaries &   & Likely binaries, consistent $v$~sin~$i$ measurements, & Section \ref{sec:binaries} & 6 & 12 & 8 \\
 & &  $v$~sin~$i$ \& $V_{eq}$ (if known) $\geq$ 11 & & & & \\
Model Comparison Sample &  & Singles, must have luminosity \& period, $V_{eq}$ $\geq$ 11 & Section \ref{sec:rad_comp} & 16 (8) & 72 (43) & 18 (7) \\  
 & Lower & 2880 $\leq$ $T_{\rm eff}$ $\leq$ 3615 K & Section \ref{sec:rad_comp} & N/A & 29 (22) & N/A \\
 & Mid  & 3616 $\leq$ $T_{\rm eff}$ $\leq$ 3999 K & Section \ref{sec:rad_comp} & N/A & 31 (13) & N/A \\
 & Upper & 4000 $\leq$ $T_{\rm eff}$ $\leq$ 4670 K & Section \ref{sec:rad_comp} & N/A & 12 (8) & N/A \\
\enddata
\tablecomments{Numbers in parentheses are the number of objects in that sample which also meet the conditions for the Resolution-Limited Sample. See Sections \ref{sec:adoption} and \ref{sec:classifying} for a detailed discussion on how the Class~II, Class~III, CTTS, and CWTTS samples were created.}
\end{deluxetable}

\subsection{Inclination Distribution} \label{sec:inc}

Several different studies have explored the distribution of stellar inclinations in open clusters, with varying results.  Some studies have suggested that clusters such as Praesepe, NGC 6791, and NGC 6819 may have some inclination alignment (e.g., \citealt{Corsaro2017, Kovacs2018}), whereas other work supports the idea that turbulence will dominate over cluster-wide rotation and result in an isotropic inclination distribution (e.g., \citealt{Menard2004, Jackson2010, Jackson2018, Jackson2019, Mosser2018, Aizawa2020}). \cite{Healy2023} examined 10 open clusters and found that 8 were consistent with isotropic spin-axis orientations and the other 2 may have a small fraction of stars that were aligned.  The expected mean inclination, $\langle$sin~$i$$\rangle$, for cluster stars with randomly distributed spin-axes is 0.785 \citep{Chandrasekhar1950}.  However, the observed $\langle$sin~$i$$\rangle$ for a sample may or may not agree with this, especially for smaller samples where individual outliers can have a large effect on the mean value. In general, we will assume an isotropic inclination distribution for the stellar rotation axes in a cluster.  For the Resolution-Limited Sample of IC~348, $\langle$sin~$i$$\rangle$ = 0.7835 $\pm$ 0.0596, in very good agreement with 0.785.  For IC~5070, the $\langle$sin~$i$$\rangle$ of the Resolution-Limited Sample is 0.6662 $\pm$ 0.0593; this is not in 2$\sigma$ agreement with 0.785, but only nine objects are used for the calculation, so this may be attributable to small number statistics.  When combined with the ONC, $\langle$sin~$i$$\rangle$ = 0.7170 $\pm$ 0.0368, which is within 2$\sigma$ agreement with 0.785.  

Under the assumption that spin-axes are randomly distributed, a $\langle$sin~$i$$\rangle$ lower than 0.785 may indicate a systematic error in one of the properties used to estimate $\langle$sin~$i$$\rangle$.  The ONC objects dominate the population of the combined sample, and \cite{Rhode2001} estimated $\langle$sin~$i$$\rangle$ = 0.64 $\pm$ 0.02.  \cite{Rhode2001} discussed potential explanations for the lower $\langle$sin~$i$$\rangle$, including errors in $v$~sin~$i$ or rotation period measurements, selection effects for stars with circumstellar disks or photometric variability, and overestimated luminosities, but eventually suggested that underestimated effective temperatures were the most likely factor.  However, as with NGC 2264 \citep{Gray2024} and IC~348, we have used the temperature scale of \cite{PM2013}, which is even slightly cooler ($\sim$50 K on average) than the temperatures used by \cite{Rhode2001}, and we did not see this effect in those other clusters.  We believe that the luminosities may be systematically overestimated due to unresolved binaries, which we had more difficulty removing from the IC~5070 and ONC samples compared to other clusters; indeed, we were able to identify several more binaries in the ONC sample than \cite{Rhode2001} had, and our average $\langle$sin~$i$$\rangle$ for the ONC sample was slightly higher than theirs at 0.727 $\pm$ 0.042.  However, without knowing the degree that the luminosity overestimation affects our sample, it is difficult to know if this is masking an overestimation of our $v$~sin~$i$ measurements. \cite{Hartmann2001} created a model for the systematic effect of unresolved binaries on luminosities, and estimated that the average log($L$) would be shifted $\sim$0.2 dex higher if all stars had an unresolved companion.  If we apply this shift to the IC~5070/ONC Resolution-Limited Sample, $\langle$sin~$i$$\rangle$ = 0.9026 $\pm$ 0.0463.  To bring $\langle$sin~$i$$\rangle$ in agreement with 0.785, we only need $\Delta$log($L$) = 0.035.  From a simple Monte Carlo simulation of a distribution similar to the one in \cite{Hartmann2001}, we find that this shift corresponds to an unresolved binary population of $\sim$20\%.


\subsection{Comparing Our $v$~sin~$i$ Measurements to the Literature}

\begin{figure}
    \centering
    \includegraphics[width=0.9\linewidth]{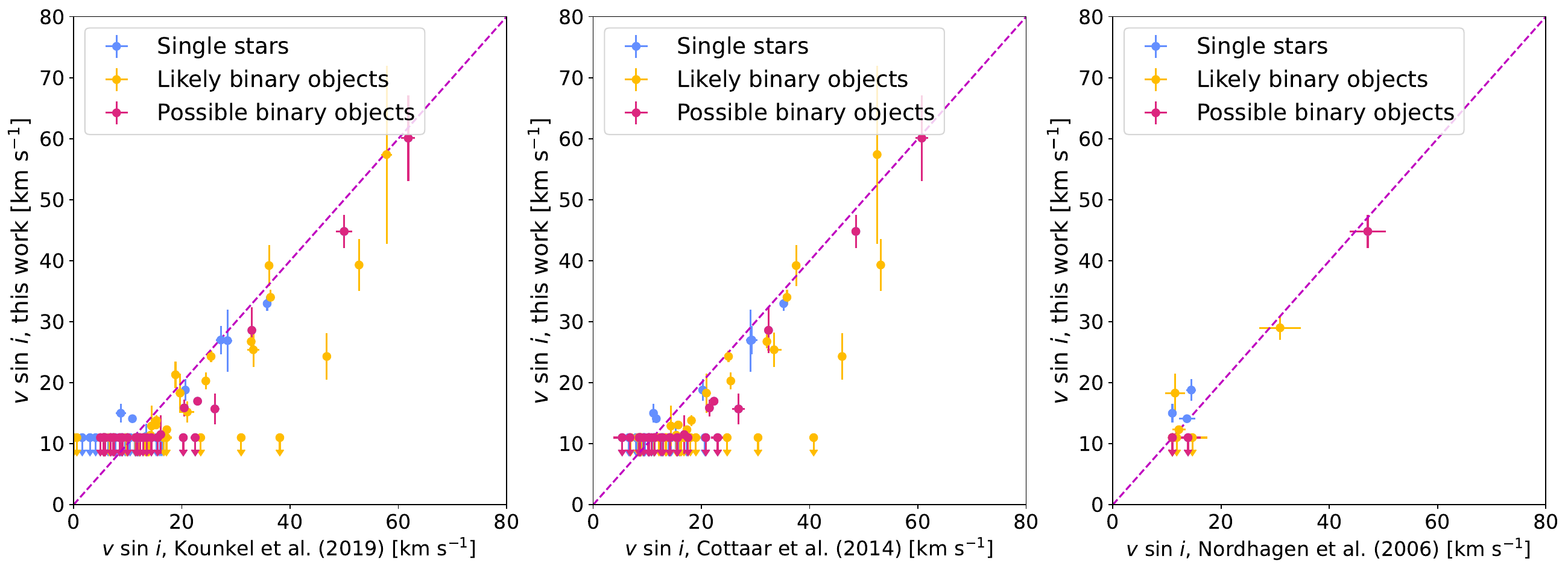}
    \caption{Comparison of our $v$~sin~$i$ measurements with measurements from \citeauthor{Kounkel2019} (\citeyear{Kounkel2019}; left), \citeauthor{Cottaar2014} (\citeyear{Cottaar2014}; center), and \citeauthor{Nordhagen2006} (\citeyear{Nordhagen2006}; right).  
    Objects with $v$~sin~$i$ $<$ 11 $\mathrm{km}\,\mathrm{s}^{-1}$ are marked as upper limits with arrows.  Objects we identified as ``likely" binary systems are marked in yellow, while ``possible" binaries are in pink, and ``single" stars are in blue.  The magenta dashed line depicts a perfect 1-to-1 correlation.  The measurements from \cite{Kounkel2019} and \cite{Cottaar2014}, which are both from APOGEE, tend to be higher than ours, although we generally find good agreement with \cite{Nordhagen2006}.}  
    \label{fig:IC348_lit_comp}
\end{figure}

One way to confirm the accuracy of our methodology and identify likely binaries is to compare our measurements to those from the literature.  Unfortunately, the stars in our IC~5070 sample do not have any previously reported $v$~sin~$i$ measurements.  A comparison between our measurements for $v$~sin~$i$ and literature values for IC~348 is shown in Fig. \ref{fig:IC348_lit_comp}.  IC~348 has $v$~sin~$i$ measurements from APOGEE reported in \cite{Cottaar2014} and \cite{Kounkel2019}.  There are 102 objects in common between our sample and \cite{Kounkel2019}, and 93 with \cite{Cottaar2014}.  The measurements from these papers are often larger than our $v$~sin~$i$ measurements; however, they are also often larger than the estimated $V_{eq}$, which is physically impossible.  For single and possible binary objects with a literature $v$~sin~$i$ and $V_{eq}$ $\geq$ 13 $\mathrm{km}\,\mathrm{s}^{-1}$ (analogous to our Resolution-Limited Sample, but using the APOGEE resolution limit and $v$~sin~$i$ measurements from \citealt{Kounkel2019}), $\langle$sin~$i$$\rangle$ = 0.905 $\pm$ 0.082.  As discussed in Section \ref{sec:inc}, this could indicate that the $v$~sin~$i$ measurements are overestimated compared to ours.  \cite{Nordhagen2006} also reports $v$~sin~$i$ for 16 objects in our IC~348 sample, and these tend to be in good agreement; for 7 objects that \cite{Nordhagen2006} measured with $v$~sin~$i$ $<$ 11 $\mathrm{km}\,\mathrm{s}^{-1}$, we also measured $v$~sin~$i$ below the resolution limit for 6 of them.  For the other nine objects with $v$~sin~$i$ $\geq$ 11 $\mathrm{km}\,\mathrm{s}^{-1}$ in \cite{Nordhagen2006}, there was $<$1$\sigma$ difference between $v$~sin~$i$ measurements for five of them, and three of the four outliers are ``likely binaries" (determined without considering the $v$~sin~$i$ measurements, i.e. by RUWE, CCF shape, literature identification).

\section{Potential Effects on $v$~sin~$i$ Distributions} \label{sec:vsini_distribs}

\subsection{The Effects of Accreting Disks on $v$~sin~$i$} \label{sec:disks}

In observations of bimodal rotation period distributions in young clusters, slow rotation seemed to correlate well with the presence of circumstellar disks (e.g., \citealt{Edwards1993, Herbst2001, Herbst2002, Lamm2005, CB2007, Venuti2017, Rebull2018}).  One proposal to explain this behavior is ``disk-locking," where the stellar rotation rate becomes synchronized to the slower rate of the inner circumstellar disk through magnetic field lines, which are accreting disk material onto the star (e.g., \citealt{Konigl1991, Shu1994a, Ostriker1995, Matt2005b, Bouvier2014}).  However, some studies have indicated no significant connection between slower rotation periods and disk-bearing stars (e.g., \citealt{Stassun1999, Rebull2001, Makidon2004, Rebull2004, Nguyen2009, Littlefair2010}).  Recently, \cite{Serna2021} and \cite{Nofi2021} have found statistically significant differences in the $v$~sin~$i$ distributions of CTTS and WTTS stars in the Orion star-forming complex and the Taurus$-$Auriga star-forming region, with WTTS stars rotating faster.  In \cite{Gray2024}, we classified stars based on a mix of IR excess and H$\alpha$ measurements from the literature: we found that a larger fraction of stars with disks and/or accretion signatures in NGC 2264 had $v$~sin~$i$ values that were below our resolution limit than those without, and that when we created a strict classification between stars with active accretion and stars with no sign of a disk, there was a statistically significant difference between the distributions at a 90\% confidence (Kolmogorov$-$Smirnov, hereafter K-S, $p$ = 0.087).  

\subsubsection{Disks in IC~5070 \& the ONC} \label{sec:disks_NAP}

\begin{figure}
    \centering
    \includegraphics[width=0.9\linewidth]{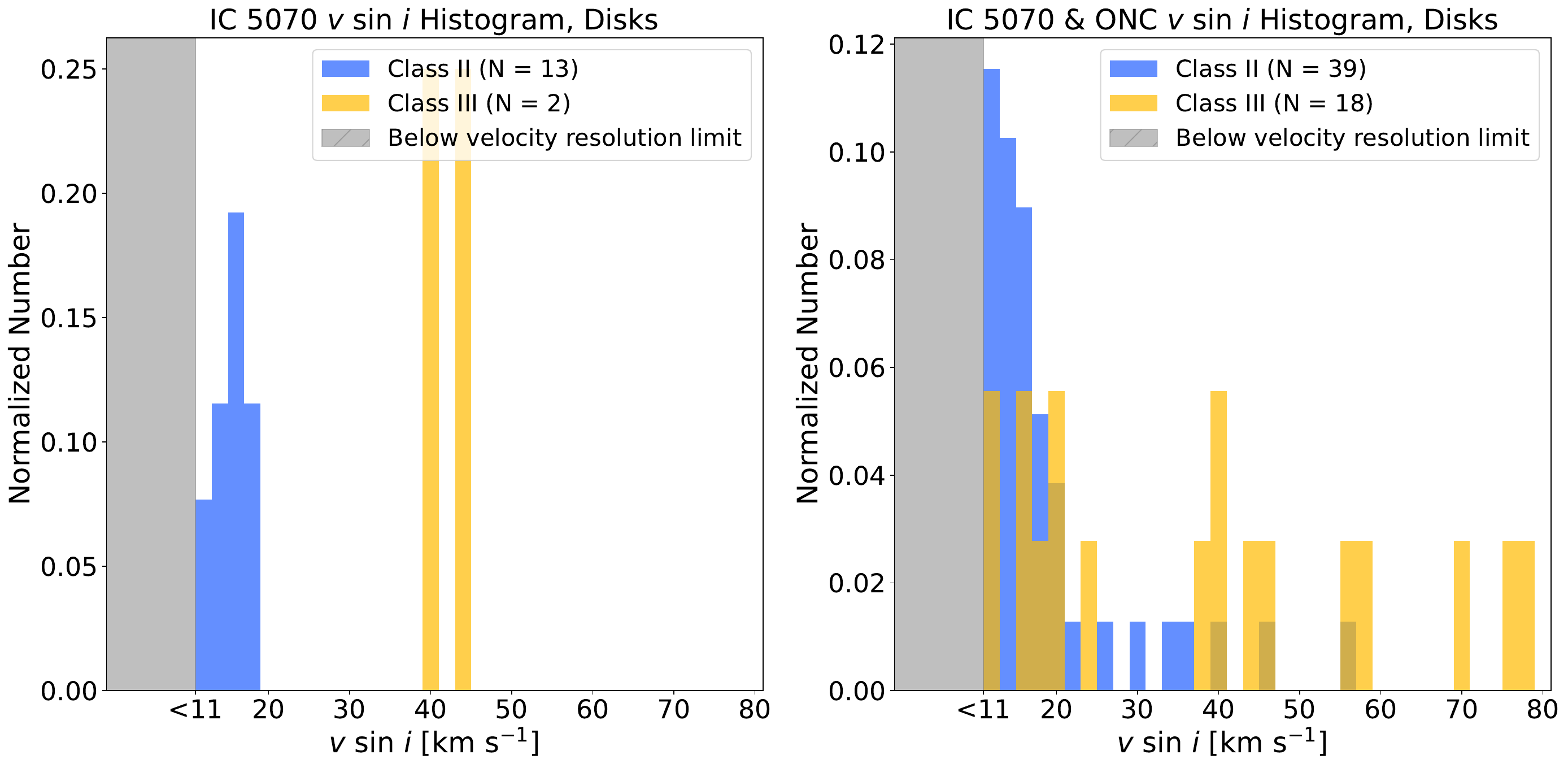}
    \caption{Normalized $v$~sin~$i$ distributions for Resolution-Limited Sample Class~II stars (blue) and Class~III stars (yellow) in IC~5070 (left) and IC~5070 combined with the ONC (right). The gray bar marks the area below the velocity resolution limit.  The histogram bins are 2 $\mathrm{km}\,\mathrm{s}^{-1}$ wide. The Class~III stars appear to rotate faster, on average, than the Class~II stars, and very few Class~II stars rotate faster than 50 $\mathrm{km}\,\mathrm{s}^{-1}$.  In a comparison of the distributions on the right, the K-S $p$-value is 0.0043, which is a statistically significant difference.} 
    \label{fig:NAP_disks}
\end{figure}

Our IC~5070 Main Analysis Sample contains 27 Class~II stars and 8 Class~III stars.  We do not have $H\alpha$ EW data from the literature for the IC~5070 stars, so we were unable to identify any CWTTSs or CTTSs using the \cite{Briceno2019} system.  The $v$~sin~$i$ distributions for the Class~II and Class~III stars in IC~5070 are shown in the left panel of Fig. \ref{fig:NAP_disks}.  Our small sample size limits the statistical analyses we are able to conduct, but the high fraction of Class~II stars does support a very young age for the cluster. 

When we increase our sample by combining IC~5070 with the ONC, we have 94 Class~II stars (19 of which are also classified as CTTS, plus another 6 CTTSs from H$\alpha$ EW alone), 28 Class~III stars, and 6 CWTTS stars.  Due to the transitional nature of CWTTSs and the small number in our sample, we exclude them from the following analysis as we wish to compare properties between two well-separated groups.  In this sample, 75\% $\pm$ 7\% (94/128) of the classified stars are Class~II, which is in line with expectations for the disked populations of clusters younger than 3 Myr \citep{Pfalzner2024}.  Errors on percentages are assumed to be Poissonian and calculated as the square root of the number of objects, normalized by the denominator.  We find that 59\% $\pm$ 8\% (55/94) of Class~II stars are below the $v$~sin~$i$ resolution limit, and 36\% $\pm$ 11\% (10/28) of Class~III stars are below the limit; these proportions are within 2$\sigma$ agreement, but not 1$\sigma$ agreement.  The right panel of Fig. \ref{fig:NAP_disks} shows the $v$~sin~$i$ distributions for the Class~II and Class~III stars in the IC~5070/ONC combined sample.  When we do a K-S test on the $v$~sin~$i$ distributions above the $v$~sin~$i$ resolution limit, we find a statistically significant difference with $p$ = 0.0043.  

However, there may be a significant portion of stars with IR excesses that are not actively accreting from their disks; \cite{Venuti2018} observed that 25.8\% of stars in their NGC 2264 sample with IR excesses did not have $H\alpha$ EWs indicative of accretion. If we limit our samples to only consider CTTSs (objects that showed accretion through $H\alpha$ EW) against Class~III objects that never showed accretion or IR excess emission, we can investigate whether the differences are likely to be due to interaction with the disk.  We have 25 CTTSs  and 28 Class~III stars, with 12 and 18 stars above the velocity resolution limit, respectively.  All CTTSs are from the ONC, as we did not have $H\alpha$ EWs for IC~5070; limiting the Class~III stars to the ONC does not change our results.  We find an even more statistically significant difference between these distributions, K-S test $p$-value = 0.0031.  When we restrict our sample to exclude stars classified as ``possible binaries", the $p$-value of the K-S test is 0.0524, which is still statistically significant at the 10\% level.

In summary, we find a statistically significant difference between the $v$~sin~$i$ distributions for Class~II and Class~III stars in our IC~5070/ONC combined sample, with Class~III stars rotating faster.

\subsubsection{Disks in IC~348} \label{sec:disks_IC348}

\begin{figure}
    \centering
    \includegraphics[width=0.5\linewidth]{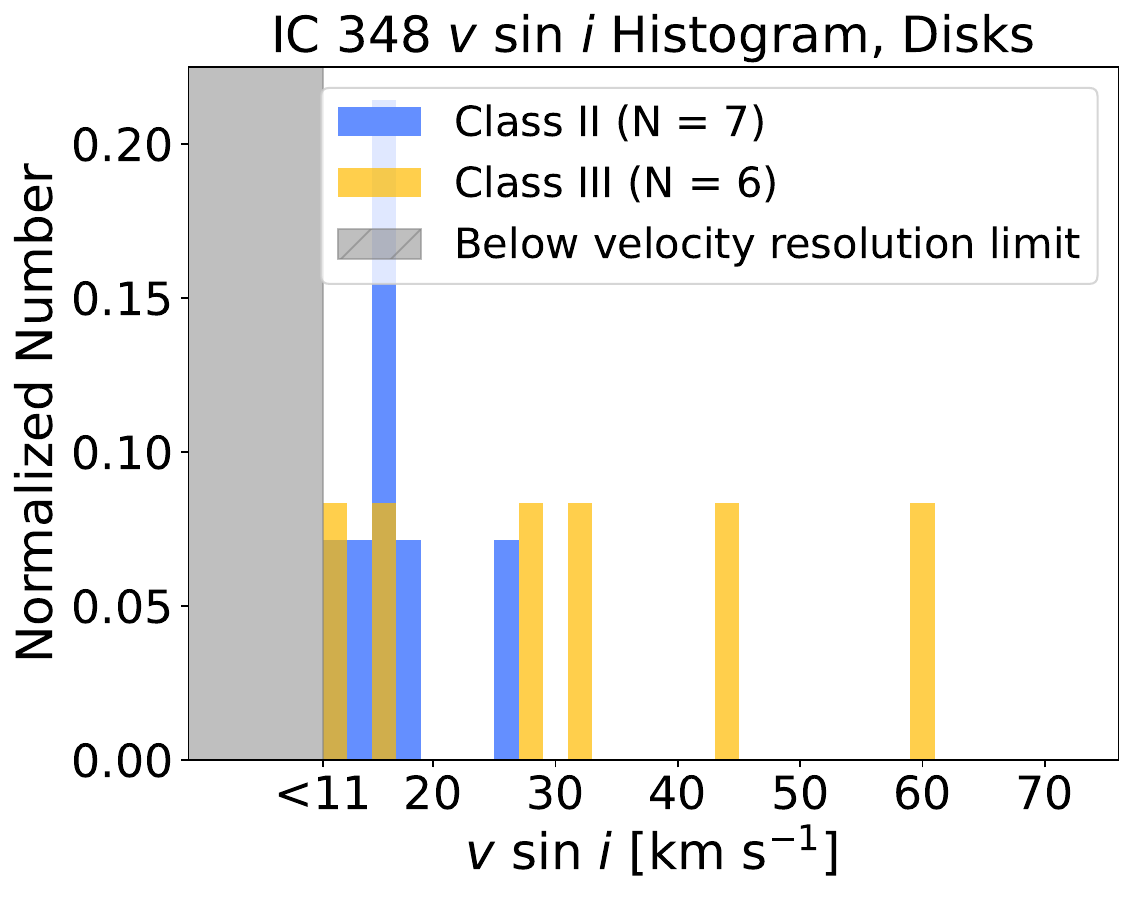}
    \caption{Normalized $v$~sin~$i$ distributions for Resolution-Limited Sample Class~II stars (blue) and Class~III stars (yellow) in IC~348. The gray bar marks the area below the velocity resolution limit.  The K-S test $p$-value = 0.091, which is statistically significant at the 10\% level.  None of the Class~II stars have $v$~sin~$i$ $>$ 27 $\mathrm{km}\,\mathrm{s}^{-1}$ while four of the Class~III stars do, though our sample size is small.}  
    \label{fig:IC348_disks}
\end{figure}

For IC~348, the sample contains 17 Class~II stars, 40 Class~III stars, and 2 CWTTSs.  This gives a disk fraction of 28\% $\pm$ 7\% (17/59), which is in agreement with estimates for other $\sim$6 Myr old clusters such as $\sigma$ Ori and Cep OB3b \citep{Hernandez2007, Allen2012}.  The $v$~sin~$i$ distribution for Class~II and Class~III stars is shown in Fig. \ref{fig:IC348_disks}.  There are seven Class~II stars and six Class~III stars above the velocity resolution limit, and the K-S test on the $v$~sin~$i$ distributions yields a $p$-value of 0.091, which is statistically significant at the 10\% level but not the 5\% level.  Our result does not change if we restrict the sample only to single stars (K-S test $p$-value = 0.095). Seven Class~II stars are also CTTSs, plus one more CTTS with no IR excess. If we compare CTTSs and Class~III stars as we did for IC~5070/ONC, the K-S test $p$-value is 0.095, although with only three CTTSs and six Class~III stars above the velocity resolution limit, this result is not definitive.  None of the Class~II stars have $v$~sin~$i$ faster than 27 $\mathrm{km}\,\mathrm{s}^{-1}$, while four of the Class~III stars do, although the sample size is small.  59\% $\pm$ 19\% (10/17) of the Class~II stars have $v$~sin~$i$ or $V_{eq}$ below the resolution limit, in very good agreement with our findings for IC~5070 and the ONC, and NGC 2264 \citep{Gray2024}, which might add some evidence to the idea that some stars with disks have their angular momentum evolution arrested or slowed by disk-locking. 

On the other hand, 85\% $\pm$ 15\% (34/40) of Class~III stars are below the resolution limit, which is unexpectedly higher than the proportion for IC~5070 and the ONC, and for NGC 2264 \citep{Gray2024}.  It is possible that this is a result of our sample size, though we do not consider this likely to be the sole explanation.  There may also be some population of stars currently classified as Class~III which have lost their disks very recently, and therefore haven't had time to spin up yet.  \cite{Serna2021} discusses how the $v$~sin~$i$ values of PMS stars may change from decreasing to increasing 
at around 5$-$6 Myr, related to the dissipation of circumstellar disks, and IC~348 appears to be an appropriate age for this phase. Stars which had disks for even a few million years appear to rotate slower than stars that never had disks, even for several million years after the disk has dissipated \citep{Serna2021}.  \cite{Kounkel2023} also finds evidence for an increased fraction of slower-rotating WTTS at higher ages, which they associate with CTTS stars transitioning to WTTS.  
However, $\sim$40\% of the Class~III stars in our IC~5070/ONC combined sample, as well as our NGC 2264 sample, were below our $v$~sin~$i$ resolution limit \citep{Gray2024}. In order to match this fraction in IC~348, at least 25 of the slower-rotating Class~III stars (representing 42\% of the total sample) must have been recently disk-locked.  This would imply that IC~348 had a substantial ($\sim$70\%) disked population at an advanced age compared to other open clusters.  Neither option seems feasible on its own, so the true explanation may be some combination of small number statistics and a recent disk loss for many stars.

To summarize our IC~348 results, for stars with $v$~sin~$i$ high enough for us to measure, we see that Class~III stars rotate faster than Class~II stars (statistically significant at the 10\% level), but our sample is fairly small.  We also find that, while the fraction of Class~II stars with $v$~sin~$i$ below our resolution limit is consistent with measurements for younger clusters ($\sim$60\%), the fraction of Class~III stars below the resolution limit is unexpectedly much higher ($\sim$85\%); we suggest that this result may be a combination of small number statistics and recent disk loss.

\subsection{The Effects of Multiplicity on $v$~sin~$i$} \label{sec:binaries}


The rotational evolution of stars in binary systems is significantly influenced by the orbital properties of the binary system (e.g., \citealt{Levato1974, Stauffer2018, Kounkel2023}).  Some studies indicate that binary companions can also truncate or disrupt disks, which would mitigate any disk-locking effects.  For example, \cite{Stauffer2018} showed that binaries in Upper Sco appeared to have shorter rotation periods than single stars in the cluster, but clarified that the slower-rotating single stars were primarily CTTSs, and the rotation period distribution of the WTTS single stars was much closer to that of the binaries (see also \citealt{Messina2019}).  \cite{Nofi2021} reported a statistically significant difference between rotation velocities of single and binary stars in the Taurus$-$Auriga star-forming region (K-S test $p$-value = 0.056); while they did not specifically compare disked single stars to disked binaries, they did conclude that interactions between a companion and the circumstellar disk may contribute to the faster average rotation velocities of the binaries.  In \cite{Gray2024}, we explored some of the relationships between disk status and binarity and found some evidence that the disked single stars were more likely to rotate slower than our velocity resolution limit than disked binaries were, while there was less of a difference between the diskless single and binary stars.  Complicating matters, \cite{Kounkel2023} identified different behaviors between very close binaries that would become tidally locked (separations $<$ a few astronomical units) and binaries that were slightly wider and less impacted by tidal interactions (separations between a few and tens of astronomical units); the very close binaries were less likely to be rapid rotators, which they attributed to being so close that they interact with their circumstellar disk in a similar way to single stars.  Meanwhile, the slightly wider binaries were more likely to be rapid rotators; this is also the separation regime where the circumstellar disks are dispersed faster for binaries than single stars \citep{Kraus2012, Kounkel2023}.  \cite{Fleming2019} investigated the timescales on which these very close binaries become tidally locked, and in their simulations, the rotation period might actually increase to approach equilibrium with the orbital period, but prior to 10 Myr, this effect is generally minimal and only for extremely close binaries ($P_{orbit}$ $<$ 10 days).  Otherwise, the binaries appeared to spin up similarly to single stars for the first $\sim$20 Myr \citep{Fleming2019}, which coincides with the findings of \cite{Kounkel2023}. 

We must take particular care when measuring the $v$~sin~$i$ of a binary object through cross-correlation; unresolved companions can artificially broaden the CCF peak through absorption line-blending, resulting in $v$~sin~$i$ being overestimated.  While double peaks are often a clear indicator of a companion, other cases may be identified by variation in $v$~sin~$i$ when measured at different times.  In our analyses, we use both our full sample of identified ``likely binaries," and a curated sample of ``quality likely binaries" selected by consistency in $v$~sin~$i$ measurements and symmetry in their CCFs.  These ``quality likely binaries" were primarily identified as binaries by differences in RV or by their RUWE.  We create a similar sample of ``quality single stars" (which by definition have consistent $v$~sin~$i$ measurements and symmetrical CCF peaks) by requiring $R_{TD}$ to be greater than or equal to the lowest $R_{TD}$ of the ``quality likely binaries."

\subsubsection{Multiplicity in IC~5070 and the ONC}

\begin{figure}
    \centering
    \includegraphics[width=0.9\linewidth]{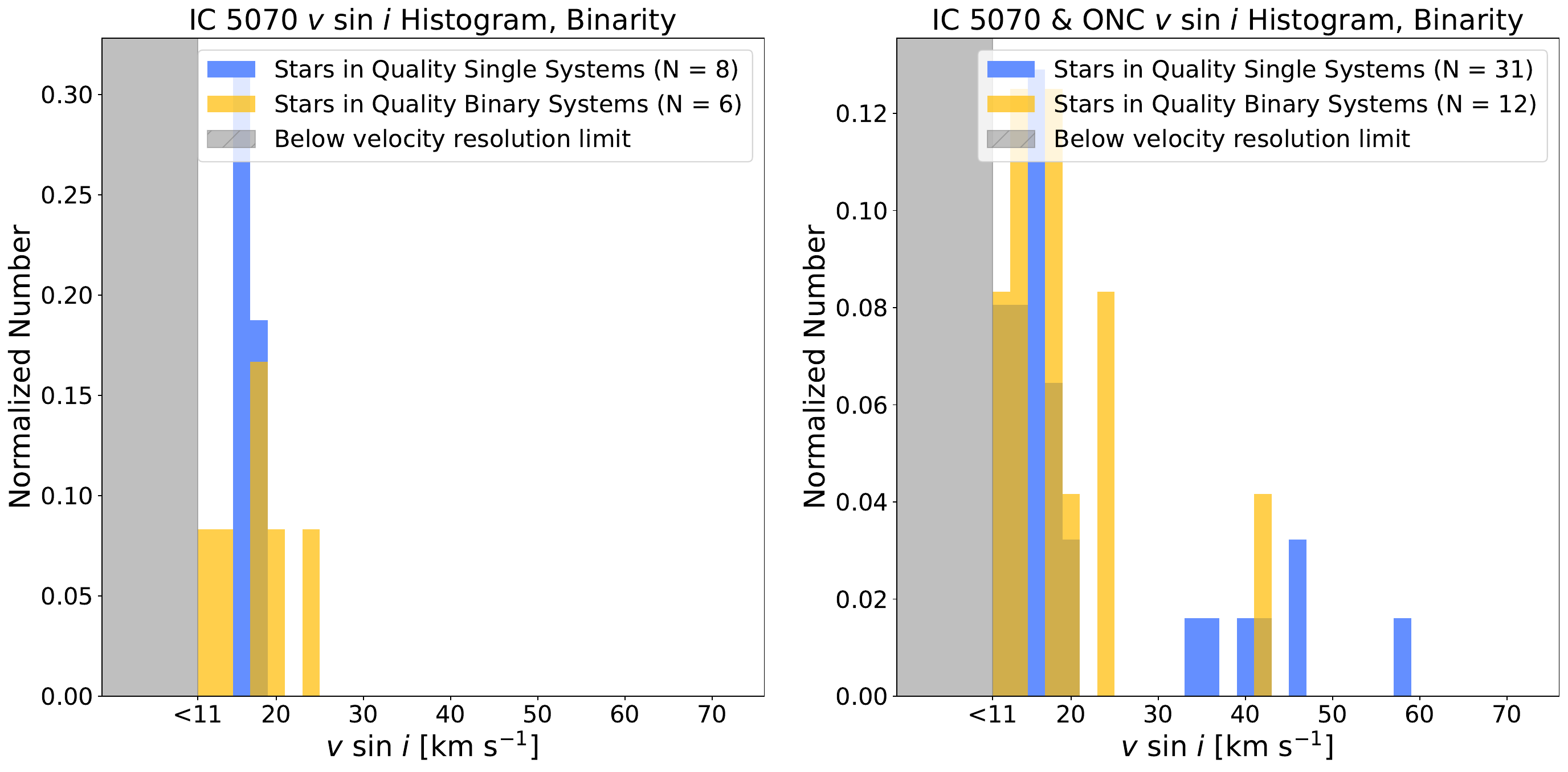}
    \caption{Normalized $v$~sin~$i$ distributions for stars we identified as ``quality single stars" (blue) and ``quality likely binary" systems (yellow) in IC~5070 (left) and IC~5070 combined with the ONC (right).  Only objects with $R_{TD}$ $\geq$ 7 are included.  The gray bar marks the area below the velocity resolution limit.  In both cases, a K-S test does not yield a statistically significant $p$-value.}  
    \label{fig:NAP_binary_distrib}
\end{figure}

For IC~5070, we have 22 single stars and 19 likely binaries.  A larger portion of the single stars have $v$~sin~$i$ or $V_{eq}$ below our resolution limit compared to the binaries: 55\% $\pm$ 16\% (12/22) of single stars versus 21\% $\pm$ 11\% (4/19) of likely binaries. The left panel of Fig. \ref{fig:NAP_binary_distrib} shows the $v$~sin~$i$ distributions for our small sample of ``quality" single and likely binary stars; we conduct our analysis using the larger sample with the ONC.   

When we combine IC~5070 with the ONC, we still see that 54\% $\pm$ 6\% (72/133) of the single stars are below the resolution limit, compared to 25\% $\pm$ 6\% (18/72) of the binary stars.  We have 12 ``quality likely binaries" and 31 ``quality single stars" ($R_{TD}$ $\geq$ 7), shown in the right panel of Fig. \ref{fig:NAP_binary_distrib}.  $\langle$sin~$i$$\rangle$ for the ``quality likely binary" sample is 0.536 $\pm$ 0.084; while we expect that this low value is a result of luminosity overestimation, this makes it difficult to tell whether we are overestimating the $v$~sin~$i$ measurements as well.  \cite{Hartmann2001} created a model for the systematic effect of unresolved binaries on luminosities, and estimated that log($L$) would be $\sim$0.2 dex higher if all stars had an unresolved companion.  If we apply this shift to our ``quality likely binary" sample, $\langle$sin~$i$$\rangle$ = 0.675 $\pm$ 0.106, which is a 1.04$\sigma$ difference from 0.785.  This lower value indicates that it is unlikely that we have systematically overestimated our binary $v$~sini~$i$ values.  For the ``quality single stars," $\langle$sin~$i$$\rangle$ = 0.733 $\pm$ 0.054, which is in agreement with 0.785.  A K-S test on the $v$~sin~$i$ distributions of these samples gives $p$ = 0.936.  Therefore, we do not observe a difference between the $v$~sin~$i$ distributions of single and binary stars at the age of the ONC and IC~5070.  

As discussed in \cite{Gray2024}, this may be because it is too early in their evolutionary history for the binary systems to have become tidally locked, or a difference in the distributions may be masked by our velocity resolution limit.  In the previously mentioned \cite{Nofi2021} study, their data had a velocity resolution limit of about 4 $\mathrm{km}\,\mathrm{s}^{-1}$; when we restricted their sample to the same limitations as ours, the K-S test $p$-value increased from 0.056 to 0.59.  This indicates that our velocity resolution limit may have a substantial impact on our results.  Additionally, with only one observation epoch for IC~5070 and no literature RVs (and similarly, only one set of RVs in the literature for the ONC), it is possible that some of the single stars would have been revealed as binaries by repeated measurements of the RV or $v$~sin~$i$ at different times, and so our samples were not perfectly separated.  However, $\langle$sin~$i$$\rangle$ for the ``quality singles" is in agreement with 0.785, so we would only expect there to be a few, if any, unresolved binaries in the sample.  

When we further split our sample by disk status, we see that, of stars with disk classifications, 80\% $\pm$ 9\% (74/93) of the single stars are classified as Class~II, while only 52\% $\pm$ 11\% (24/46) of the binary stars are Class~II.  This may indicate that binary systems are less likely to have disks (and therefore be disk-locked) than single stars.  We find that 57\% $\pm$ 9\% (42/74) of Class~II single stars have $v$~sin~$i$ below the resolution limit, but only 21\% $\pm$ 9\% (5/24) of Class~II binaries do, which is in very good agreement with our findings for NGC 2264 \citep{Gray2024}.  We find similar percentages for the Class~III stars (47\% $\pm$ 16\% for the singles and 27\% $\pm$ 11\% for the binaries); if disk interaction and not binarity was the main driver for the difference, we would expect the fractions for the Class~III stars to be more similar to each other.  We note that the difference between the percentages for Class~II stars are 2.83$\sigma$ and only 1.03$\sigma$ for the Class~III stars.  However, as the young age of the stars in this sample means we have a smaller number of Class~III stars (19 single Class~III stars and 22 binary Class~III stars), we cannot draw a firm conclusion.

We summarize our findings for our IC~5070/ONC combined sample as follows: We do not find a statistically significant difference between the $v$~sin~$i$ distributions for single and binary stars, which we suggest may be due to the young age of the stars in the sample or our velocity resolution limit restricting our comparison of the slower ends of the distributions.  We suggest that single stars with disks may be more likely to be slow rotators than binary stars with disks, and that this difference might be less significant between single stars and binaries without disks, although we emphasize that our sample sizes are small.

\subsubsection{Multiplicity in IC~348}

\begin{figure}
    \centering
    \includegraphics[width=0.5\linewidth]{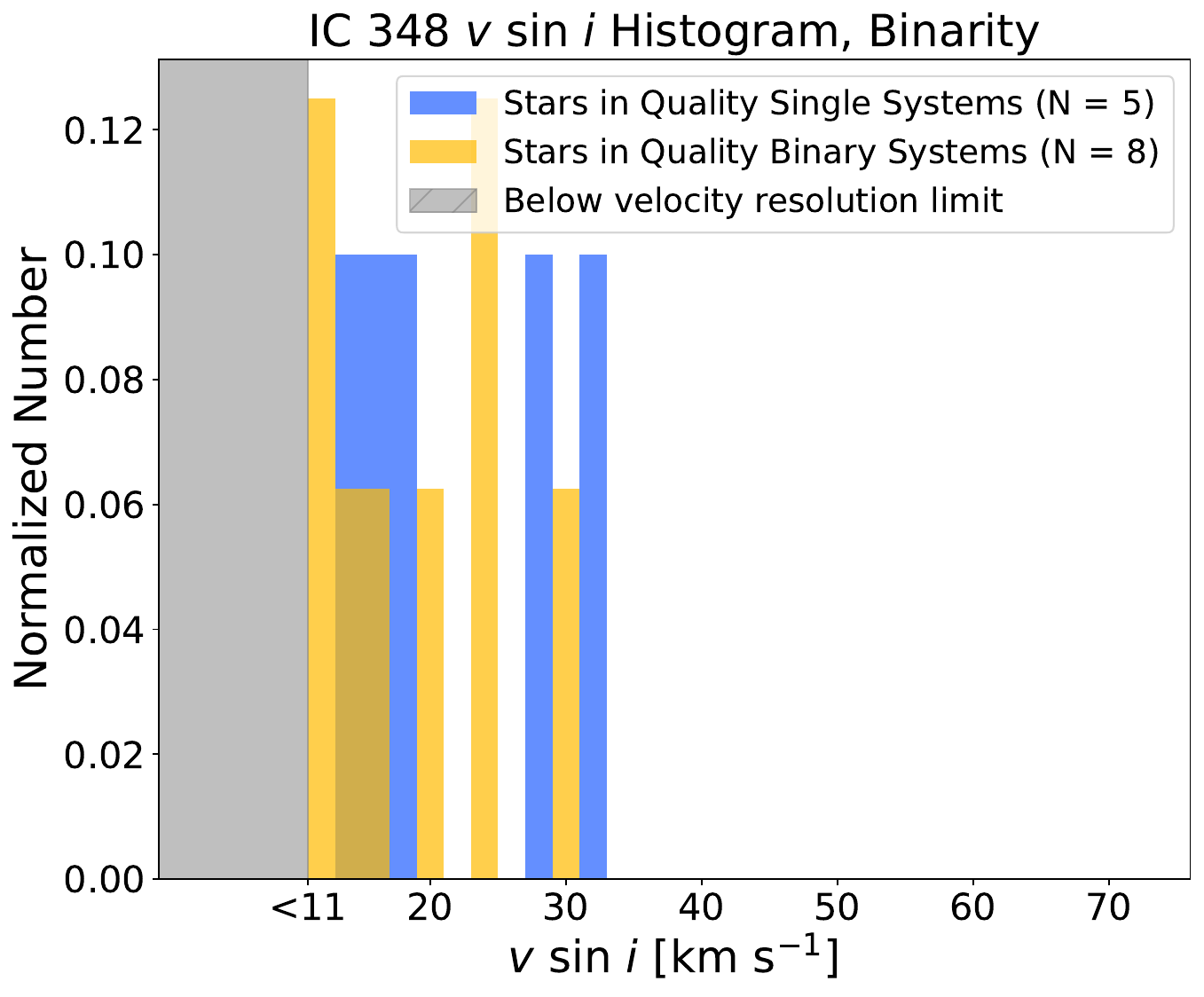}
    \caption{Normalized $v$~sin~$i$ distributions for stars we identified as ``quality single stars" (blue) and ``quality likely binary" systems (yellow) in IC~348.  Only objects with $R_{TD}$ $\geq$ 4.4 are included.  The gray bar marks the area below the velocity resolution limit. A K-S test does not yield a statistically significant $p$-value.}  
    \label{fig:IC348_binary_distrib}
\end{figure}

For IC~348, 76\% $\pm$ 16\% (22/29) single objects have $v$~sin~$i$ or $V_{eq}$ below the velocity resolution limit, while only 51\% $\pm$ 11\% (20/39) binaries do.  We observed a larger fraction of slower rotators in IC~348 overall compared to IC~5070, the ONC, and NGC 2264, which were all fairly consistent with each other.  Fig. \ref{fig:IC348_binary_distrib} shows our 8 ``quality likely binaries" and 5 ``quality single stars" ($R_{TD}$ $>$ 4.4); as our sample is small, we present the following analysis without drawing firm conclusions.  First, we must again check whether we are systematically overestimating the $v$~sin~$i$ for the ``quality likely binaries."  For this sample, $\langle$sin~$i$$\rangle$ = 0.704 $\pm$ 0.053.  Again, we expect that overestimated luminosities affect $\langle$sin~$i$$\rangle$; a log($L$) shift of 0.2 dex gives $\langle$sin~$i$$\rangle$ = 0.887 $\pm$ 0.067, which may indicate that the $v$~sin~$i$ values are overestimated.  However, the small size of the sample makes $\langle$sin~$i$$\rangle$ very sensitive to any errors in rotation periods.  As an example of how a different period can affect the average, one star has a period in the literature that is half of our adopted value; if we use this, $\langle$sin~$i$$\rangle$ = 0.647 $\pm$ 0.059, and after the luminosity adjustment, $\langle$sin~$i$$\rangle$ = 0.814 $\pm$ 0.064, which would be in agreement with 0.785.  For the ``quality single stars", $\langle$sin~$i$$\rangle$ = 0.835 $\pm$ 0.056, which is also in agreement with 0.785 as expected.  When we compare the $v$~sin~$i$ distributions of the quality single and binary stars with a K-S test, $p$ = 0.684, which is not a statistically significant difference.  We had measurements from different epochs for most of our objects in IC~348, and in combination with RUWE and careful examination of the CCF peaks, we are confident that the similarity is not due to poor separation between the samples.  However, we do have a very small sample size and the large fraction of stars below the velocity resolution limit impacts our ability to evaluate the full $v$~sin~$i$ distribution, so it is difficult for us to draw a conclusion from this result.  

In summary, we do not find a statistically significant difference between the $v$~sin~$i$ distributions for single and binary stars in IC~348; while we were confident in our ability to separate single and binary stars, our analysis is limited by our small sample size and $v$~sin~$i$ resolution limit.

\section{Statistical Comparison of Projected Stellar Radii to Radii Predicted by Stellar Evolutionary Models} \label{sec:rad_comp}

\subsection{Maximum Likelihood Statistical Analysis Method} \label{sec:maxlikelihoodmethod}

We perform a statistical study of the projected stellar radii for our samples in comparison to radii predicted by stellar evolution models, to investigate ``radius inflation" for different model ages and starspot coverages.  One nuance of the ``radius inflation" problem is that the degree of inflation can be very sensitive to the age of the model used, particularly for stars in very young clusters.  For NGC 2264, which has age estimates in the literature ranging from 1 to 5 Myr (e.g., \citealt{Sung1997, Flaccomio2000, Venuti2018}), the estimated radius inflation with the \cite{Baraffe2015} models ranged from 1\% at 1 Myr to 31\% at 5 Myr \citep{Gray2024}.  Assuming that the ages of the stars in the sample are representative of the cluster's age, the ``radius inflation" occurs because the effective temperature predicted by the starspot-free model at a given luminosity and age is too hot, which causes the model to underestimate the radius at the correct age of the cluster.  
This is well illustrated by a study of the PMS eclipsing binary UScoCTIO 5 by \cite{Kraus2015}, which showed that the \cite{Baraffe2015} model successfully predicted the masses of the components (based on luminosity and age estimates of the high-mass MS stars in the cluster) but underestimated the radii.  At the correct mass of the stellar components, all effective temperatures predicted by the \cite{Baraffe2015} models were too hot no matter which age was used \citep{Kraus2015}.

Two mechanisms that would reduce the effective temperature of the stars have been proposed, both linked to the high magnetic activity of TTSs: inhibition of energy transfer through convection, and increased starspot coverage on the stellar surface (e.g., \citealt{Gough1966, Feiden2013, Somers2015a, Somers2020, Kiman2024}). \cite{Jackson2016} and \cite{Jackson2018} demonstrated that magnetic inhibition of convection is less effective than starspots in explaining radius inflation for fully convective, low-mass stars, which describes the sample of stars in this work. Therefore, we focus on the potential impacts of starspots by using the 
SPOTS models of \cite{Somers2020}, which include six starspot coverages from 0\% to 85\%.  
Starspots reduce the effective temperature of the star, but do not significantly reduce the luminosity, which leads to a larger estimated radius \citep{Somers2020}.  \cite{Somers2020} demonstrated that a modest starspot coverage of 34\% accurately predicted the temperature and radius of the components of UScoCTIO 5 at the age of the host cluster without affecting the predicted luminosity or mass.

Recent work in the Taurus$-$Auriga star-forming complex has suggested that large fractions of the surfaces of TTSs may be covered in starspots.  \cite{Gangi2022} characterized over a quarter of their CTTS sample and two-thirds of their WTTS sample as ``heavily spotted," with an average of 40$-$50\% of the total flux coming from cool regions on the stellar surface.  \cite{PerezPaolino2024} modeled spectra for 10 WTTS stars and found starspot fractions between 42\% and 84\% for all of them, with an average of 66\%; in a follow-up study, they modeled 16 CTTS stars and found an average starspot coverage of 50$-$60\% \citep{PerezPaolino2025}. Conversely, \cite{Herbert2023} used multifilter photometric variability to estimate that the stellar surfaces of YSOs in IC~5070 were usually less than 20\% covered in starspots.  One possible explanation for the difference between these estimates might be that the simple magnetic fields of the convective PMS stars cause more starspots to form at higher latitudes \citep{PerezPaolino2024} The photometric variability of the star due to rotation could be smaller for more inclined stars, which could result in an underestimated spot coverage \citep{Herbert2023, PerezPaolino2024}.  At the same time, the cool regions could cover a large portion of the visible area of the star for an inclined star and be consistently visible in the stellar spectra, leading to an overestimated coverage fraction \citep{PerezPaolino2024}. 

We compare both starspot-free \citep{Baraffe2015} and starspot \citep{Somers2020} models to our projected radius estimates in our statistical analysis.  Both models assume solar metallicity, are nonaccreting, and are appropriate for PMS stars.  We follow the method to create a maximum likelihood model from \cite{Jackson2018}, described in detail in \cite{Gray2024}.  In summary, for each star, we measure the ratio between the projected stellar radius $R$~sin~$i$ = $P$ $v$~sin~$i$/(2$\pi$) and the radius predicted by the model, $R_m$, and generate a probability distribution of possible ratios using the uncertainty in the $v$~sin~$i$ measurement, a given distribution of inclinations, and an average of the radius ratio, $\rho$ = $R$/$R_m$.  We then find the value of $\rho$ where the log-likelihood function summed over all stars, ln~$\widehat{\mathcal{L}}$, is maximized, and estimate the uncertainty of $\rho$ from the standard deviation of the likelihood function distribution.


As discussed in Section \ref{sec:inc}, the observed $\langle$sin~$i$$\rangle$ for a sample may or may not agree with 0.785, especially for smaller samples where individual outliers can have a large effect on the mean value. The quantity $\rho$ is the average ratio of the true radius to model radius, measured from a distribution of $R$~sin~$i$/$R_m$; an observed distribution where $\langle$sin~$i$$\rangle$ is higher or lower than the expected 0.785 could then lead to an overestimated or underestimated $\rho$.  To correct for this, while we start by assuming a random distribution of spin-axes, we also use Equation 6 in \cite{Jackson2010} to modify the cone angle ($\lambda$) and inclination with respect to the observer ($\alpha$) to change the assumed $\langle$sin~$i$$\rangle$ of the model to match that of the observed distribution.   
We select the largest $\lambda$ (weakest alignment) and corresponding $\alpha$ that gives a $\langle$sin~$i$$\rangle$ within 0.001 of the observed $\langle$sin~$i$$\rangle$.  However, we do not specifically model the distribution, so we do not consider these parameter selections to be any prediction about the true distribution of the cluster.  We report $\rho$ using both the assumed random distribution and the adjusted distribution.

\subsection{Comparing Measured Radii to Stellar Evolution Model Predictions} \label{sec:rho_results}


We present the results of our maximum likelihood method comparing our geometric radii to radii predicted by two sets of stellar evolution models, the starspot-free \cite{Baraffe2015} models and the variable starspot coverage \cite{Somers2020} models.  In this statistical analysis, $\rho$ = 1 indicates that a model is a good fit to the observed radii, while $\rho$ $>$ 1 indicates radius inflation, that the measured radii tend to be larger on average than the radii predicted by the model.  Overall, we look at the predictions of the models at different ages and different starspot coverages. In general, younger models will have lower $\rho$ than comparable older models.  For models that are the same age, those with higher fractions of starspot coverage will have lower $\rho$ compared to the starspot-free models.  This can lead to a model with high starspot coverage and a younger model with no starspots estimating the same $\rho$, with consequences for the age estimation of the stars in the sample and the cluster at large. 
TTSs are expected to have a moderate to high fraction of starspots ($>$50\%; \citealt{Stauffer2003, Gangi2022, PerezPaolino2024}), and we consider this when we compare the $\rho$ values to evaluate which age and starspot coverage model is the most appropriate for a sample.  

The stellar radii are contracting quickly, so the fit of the model is fairly sensitive to the age of the stars.  Some studies have observed a mass-dependence in age estimation for very young clusters, where higher-mass stars appear to have older ages than lower-mass stars in the same cluster (e.g., \citealt{Palla2000, Hartmann2003}), so we compare the model radius predictions to the entire Model Comparison Sample (``all" objects) as well as to three temperature-based subgroups.  As the stars in these clusters are still on the Hayashi track \citep{Hayashi1961}, temperature is more consistent than luminosity at the estimated age ranges, so it is a good proxy for mass.  \cite{Somers2020} suggested that models with higher starspot fractions may be sufficient to close or at least minimize the gap between higher- and lower-mass stars. 

\begin{figure}
    \centering
    \includegraphics[width=0.9\linewidth]{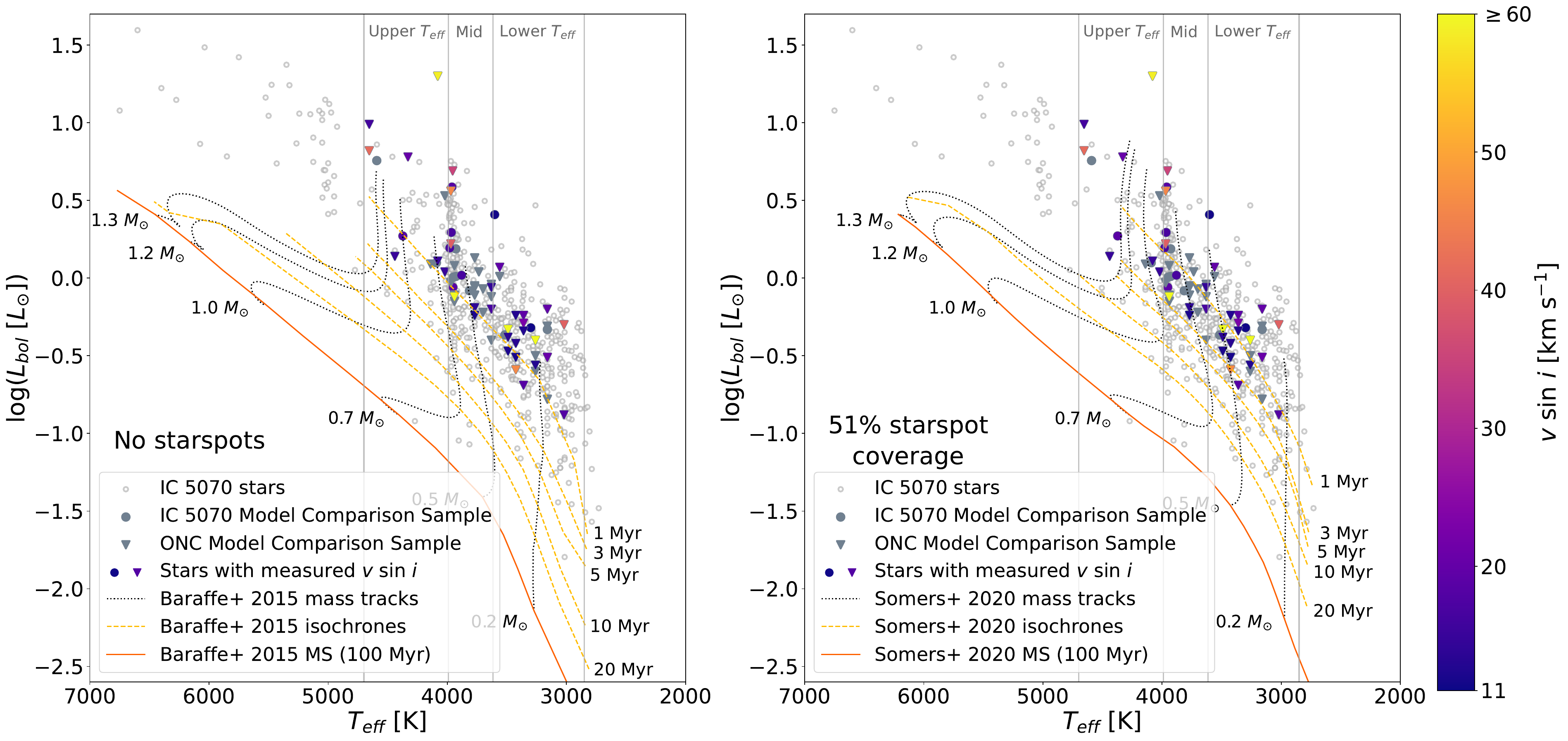}
    \caption{H-R diagrams showing isochrones (dashed yellow lines) and mass tracks (dotted black lines) from the starspot-free \cite{Baraffe2015} model (left) and the 51\% starspot coverage \cite{Somers2020} model (right).  The main-sequence track for both models is approximated by the 100 Myr isochrone (solid orange line).  The Model Comparison Sample for IC~5070 (circles) combined with the ONC (downward-pointing triangles) is plotted with filled symbols; the color of the markers indicates the measured $v$~sin~$i$ of the stars, with gray shapes for stars with $v$~sin~$i$ or $V_{eq}$ below the 11 $\mathrm{km}\,\mathrm{s}^{-1}$ velocity resolution limit.  Stars from our full catalog of PMS stars in IC~5070 are plotted as light gray open circles for comparison, showing that our sample is representative of where most of these stars are located in the H-R diagram.  The vertical gray lines mark the main temperature-based subgroups that were also used in analysis for the combined sample.  The ONC and IC~5070 occupy the same regions on the H-R diagrams, indicating that they are roughly the same age and can be combined for analysis related to evolutionary stages.}  
    \label{fig:NAP_HR_diagrams}
\end{figure}

\begin{figure}
    \centering
    \includegraphics[width=0.9\linewidth]{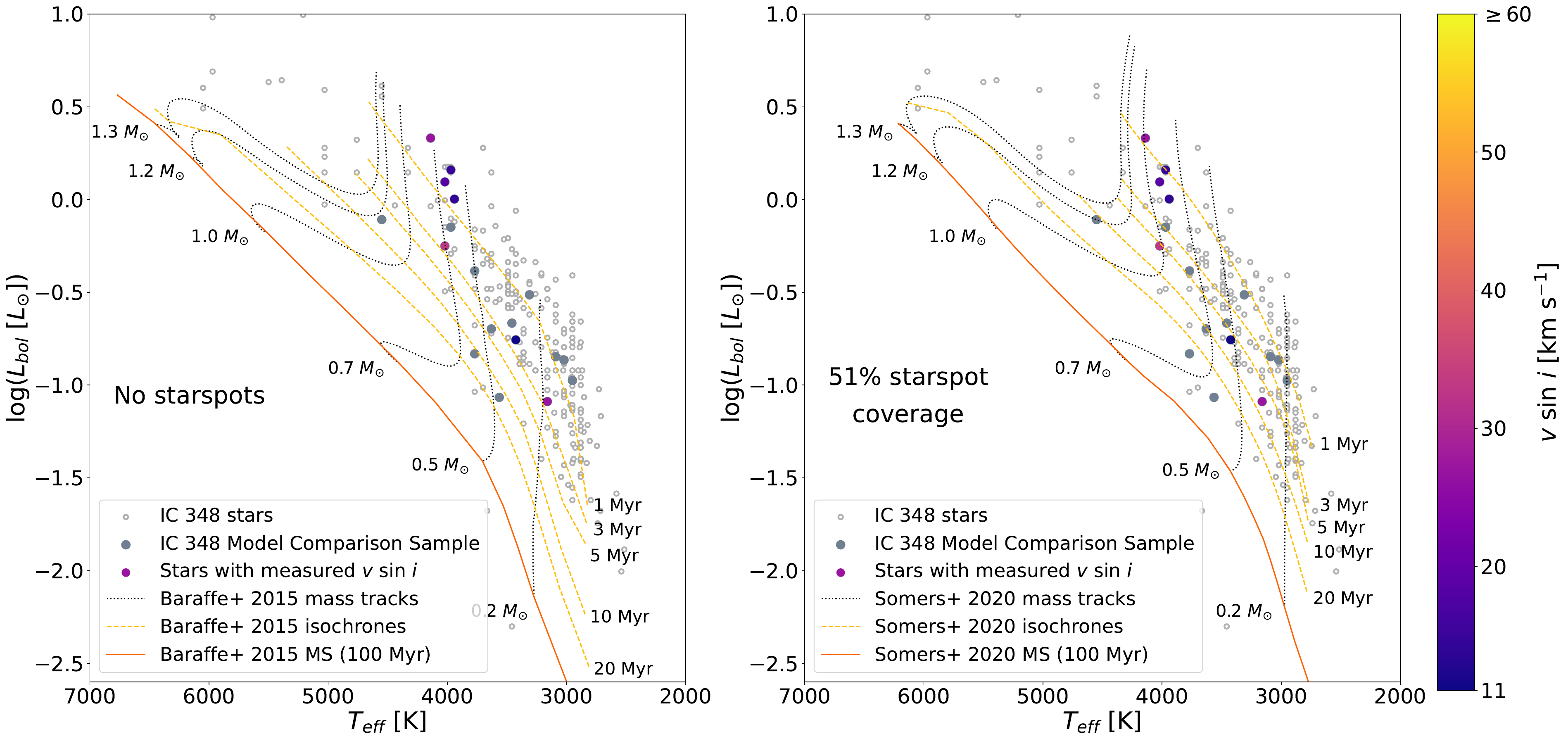}
    \caption{H-R diagrams showing isochrones (dashed yellow lines) and mass tracks (dotted black lines) from the starspot-free \cite{Baraffe2015} model (left) and the 51\% starspot coverage \cite{Somers2020} model (right).  The main-sequence track for both models is approximated by the 100 Myr isochrone (solid orange line).  The Model Comparison Sample for IC~348 is plotted in filled circles, with the color indicating the measured $v$~sin~$i$ of the stars; stars with $v$~sin~$i$ or $V_{eq}$ below the 11 $\mathrm{km}\,\mathrm{s}^{-1}$ velocity resolution limit are represented by gray circles.  Stars from our full catalog of PMS stars in IC~348 are plotted as light gray open circles for comparison, showing that our sample is in the same region of the H-R diagram as most of these stars.}  
    \label{fig:IC348_HR_diagrams}
\end{figure}

In \cite{Gray2024}, we discussed that the observed age difference across masses may be due to differences in birthline effects for higher-mass stars, as proposed by \cite{Hartmann2003}, \cite{Hartmann2016} and \cite{Hosokawa2011}, so we define the ``upper" temperature group as stars with $T_{\rm eff}$ $\geq$ 4000 K.  At the age of IC~5070, this roughly correlates with stars $\gtrsim$0.6 $M_{\odot}$ in the \cite{Baraffe2015} models and the 0\% starspot coverage version of the \cite{Somers2020} models.  We divide the remaining stars equally into the ``lower" and ``mid" groups.  For IC~5070, we only have 16 stars in the Model Comparison Sample, 8 of which have a measured value of $r$~sin~$i$ ($v$~sin~$i$ $\geq$ 11 $\mathrm{km}\,\mathrm{s}^{-1}$).  However, when we add the stars from the ONC, we increase the sample size enough to split it into subgroups.  For the IC~5070/ONC combined sample, the ``all" group contains 72 stars, 43 of which have a measured value of $r$~sin~$i$, the ``lower" group has 29 stars (22 with $r$~sin~$i$) and temperatures ranging from 2880 to 3615 K, the ``mid" group has 31 stars (13 with $r$~sin~$i$) and temperatures ranging from 3616 to 3999 K, and the ``upper" group has 12 stars (8 with $r$~sin~$i$) with temperatures between 4000 and 4670 K. Using the two starspot-free models, the division between the ``lower" and ``mid" groups occurs at $\sim$0.4 $M_{\odot}$.  Using models with increasing starspot coverages increases the estimated mass ranges for each temperature subgroup, with each group boundary increasing roughly $\sim$0.3$-$0.4 $M_{\odot}$ from 0\% to 85\% coverage.  IC~348 has 18 stars in the Model Comparison Sample, 7 of which have measured $r$~sin~$i$ values; as with IC~5070, the sample is not large enough to split into subgroups.  The H-R diagrams illustrate the Model Comparison Samples and temperature subgroups in Figures \ref{fig:NAP_HR_diagrams} (IC~5070/ONC) and \ref{fig:IC348_HR_diagrams} (IC~348), relative to the starspot-free \cite{Baraffe2015} and 51\% starspot coverage \cite{Somers2020} models.  The stars from our complete cross-matched catalogs assembled in Section \ref{sec:targets} are plotted in the background to show that our Model Comparison Samples are representative of where PMS stars in these clusters are located on the H-R diagram and can be considered the same age as the larger cluster.  There is a significant overlap between IC~5070 and ONC stars on the H-R diagram, further indicating that they are very close to the same age.  

While we report the $\rho$ values derived from assuming a random, isotropic orientation of spin-axes, we also report $\rho$ calculated using the observed $\langle$sin~$i$$\rangle$ for the samples.  Assuming a $\langle$sin~$i$$\rangle$ that is higher than what is observed for the sample can lead to an underestimated $\rho$, which would result in an overestimated age.  Additionally, because $\langle$sin~$i$$\rangle$ is slightly different for different subgroups, adjusting the maximum likelihood model to use the observed $\langle$sin~$i$$\rangle$ for the sample mitigates the effects of the sample $\langle$sin~$i$$\rangle$ on the average $\rho$.  For these reasons, we base our conclusions on the $\rho$ values calculated with the adjusted $\langle$sin~$i$$\rangle$.  For IC~5070, $\langle$sin~$i$$\rangle$ is 0.679.  For the IC~5070/ONC sample, $\langle$sin~$i$$\rangle$ for the ``all" group is 0.694 $\pm$ 0.037, for ``lower," it is 0.649 $\pm$ 0.047, for ``mid," it is 0.726 $\pm$ 0.071, and for ``upper," it is 0.767 $\pm$ 0.091.  For IC~348, $\langle$sin~$i$$\rangle$ is 0.732.

\subsubsection{IC~5070 \& the ONC}

\begin{figure}
    \centering
    \includegraphics[width=0.9\linewidth]{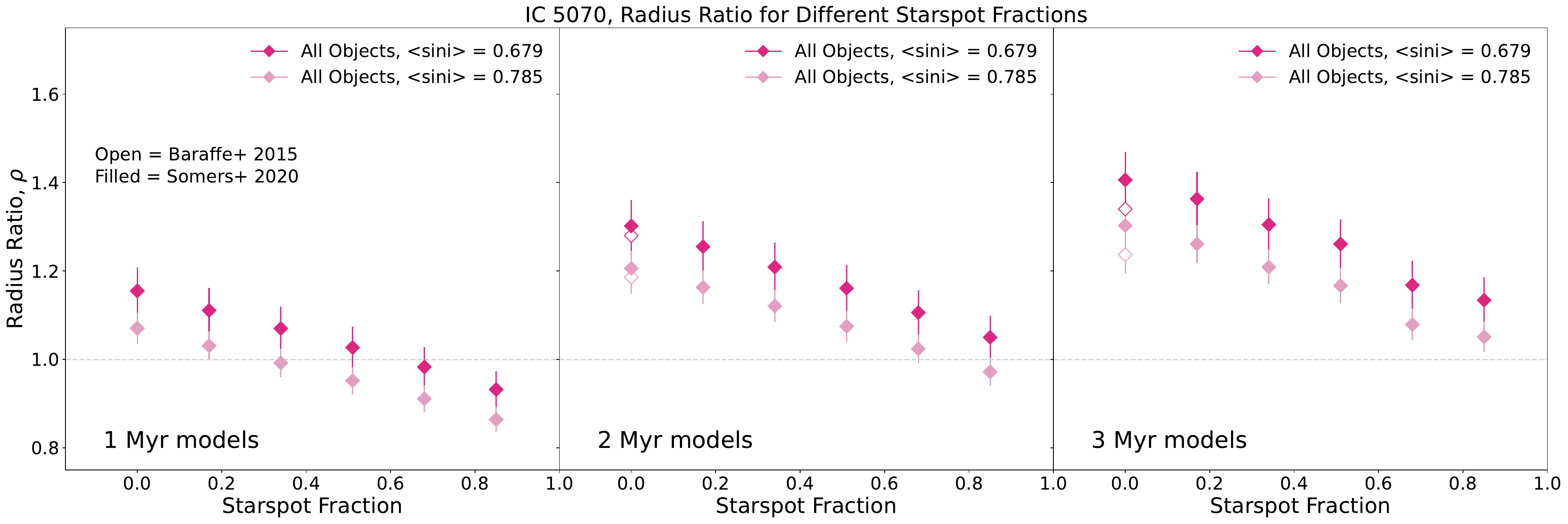}
    \includegraphics[width=0.9\linewidth]{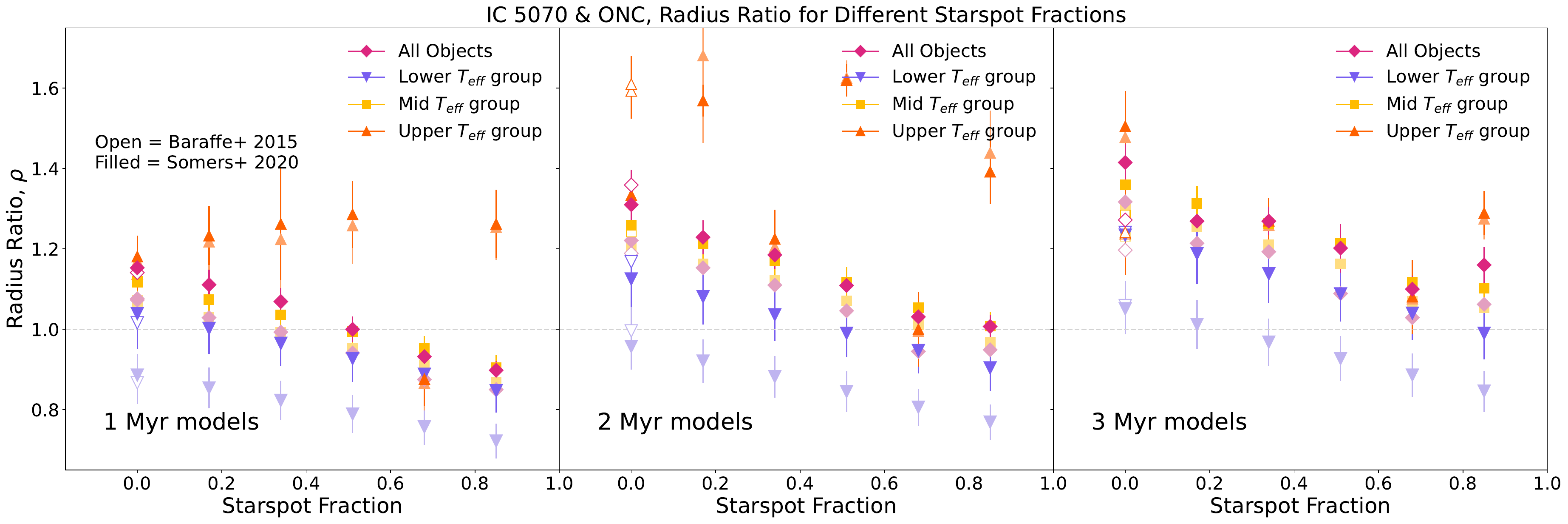}
    \caption{Top: IC~5070.  Bottom: IC~5070 and ONC.  Radius ratio of radius from $v$~sin~$i$ and period measurements to radius predictions from models with different starspot coverage fractions, assuming ages of 1 Myr (left), 2 Myr (center), and 3 Myr (right).  The \cite{Baraffe2015} model (open symbols) only appears at 0\% starspot fraction because that model does not include starspots, and the \cite{Somers2020} models use filled symbols.  The lighter symbols indicate that $\rho$ was calculated assuming an isotropic inclination distribution, $\langle$sin~$i$$\rangle$ = 0.785, and the darker symbols are $\rho$ values that have been calculated with $\langle$sin~$i$$\rangle$ adjusted to match the input sample. For IC~5070, radius inflation is still apparent at 3 Myr even with 85\% starspot coverage, indicating that 1$-$2 Myr may be a more appropriate age estimate.  The combined sample is large enough to divide objects by mass, using effective temperature as a proxy: all objects (pink diamonds), lower $T_{\rm eff}$ group (purple downward-pointing triangles), mid $T_{\rm eff}$ group (yellow squares), and upper $T_{\rm eff}$ group (orange upward-pointing triangles).  We find 2$\sigma$ agreement between the lower and mid groups, indicating that there is not a difference in the age estimates for stars in those mass ranges, but the upper group is too erratic for meaningful analysis.} 
    \label{fig:NAP_rsini_rho}
\end{figure}

Figure \ref{fig:NAP_rsini_rho} shows the average radius ratio, $\rho$, for IC~5070 (top) and the IC~5070/ONC combined sample (bottom), and these values are tabulated in Tables \ref{tab:NAP_Baraffe_rho}$-$\ref{tab:NAP-ONC_SPOTS_rho}, located in the Appendix.  If we first consider the IC~5070 Model Comparison Sample alone, we see that while starspots reduce the radius inflation, at 3 Myr, even 85\% starspot coverage does not fully address the problem, so the stars appear to be younger than 3 Myr.  Similarly, an average coverage of 85\% starspot coverage is required for the radius inflation to be fully minimized ($\rho$ $\simeq$ 1) at 2 Myr, so this appears to be an upper age limit for the cluster.  We still see some radius inflation at 1 Myr for the starspot-free models, indicating that the cluster may be slightly younger than this, but an average starspot coverage of 51\% is enough to avoid radius inflation.  Therefore, we estimate that the stars in our sample are likely between 1 and 2 Myr. 

We did a similar analysis for the ONC sample on its own and found that it might be slightly older, with $\rho$ up to 20\% lower than IC~5070 for some models, but generally averaging only 8\% lower and still fitting best between 1$-$2 Myr, so we can combine them to increase our sample size for stars in this stage of development in order to compare results for mass subgroups.  For this portion of the analysis, the focus is on the relative $\rho$ between mass subgroups rather than age estimation.  
In the combined IC~5070/ONC sample, once we have compensated for the different values of $\langle$sin~$i$$\rangle$, the ``lower" and ``mid" groups have $\rho$ within 2$\sigma$ agreement.  While we emphasize that the subgroups contain a fairly small number of stars with defined $v$~sin~$i$ measurements, this provides evidence that the apparent difference in age estimates at different masses is not observed for stars with temperatures below 4000 K.  At a maximum starspot coverage of 85\%, the 1$-$2 Myr models predict stars at these temperatures to have masses below 1.1 $M_{\odot}$.  We also did not see a reduction in the magnitudes of the differences between the two subgroups at increasing starspot coverages.  \cite{Somers2020} explored this effect across a larger mass range of stars, which did have a statistically significant difference in age estimates without starspots, so it may be that the stars in our sample were a similar enough mass that the model affected them to the same degree.  The ``upper" group appears to be more inflated, which would align with a slightly younger age estimate compared to the ``lower" and ``mid" groups.  However, in contrast to the consistent behavior of $\rho$ for other subgroups between models of different ages and starspot coverages, $\rho$ for the ``upper" group is very erratic; we attribute this to the small size of the ``upper" group and do not attempt to draw further conclusions.  

In summary, we estimate that the stars in our sample from IC~5070 are $<$2 Myr old, but can be fit to a 1 Myr model with a starspot coverage $\geq$50\%.  We see a 2$\sigma$ agreement in relative estimated radius inflation for stars with temperatures $<$ 4000 K, so we do not find evidence for a mass-dependence in age estimation in that associated range of masses, but we are unable to make comparisons to our higher-mass stars because the results did not follow a consistent trend between models.

\subsubsection{IC~348}

\begin{figure}
    \centering
    \includegraphics[width=0.9\linewidth]{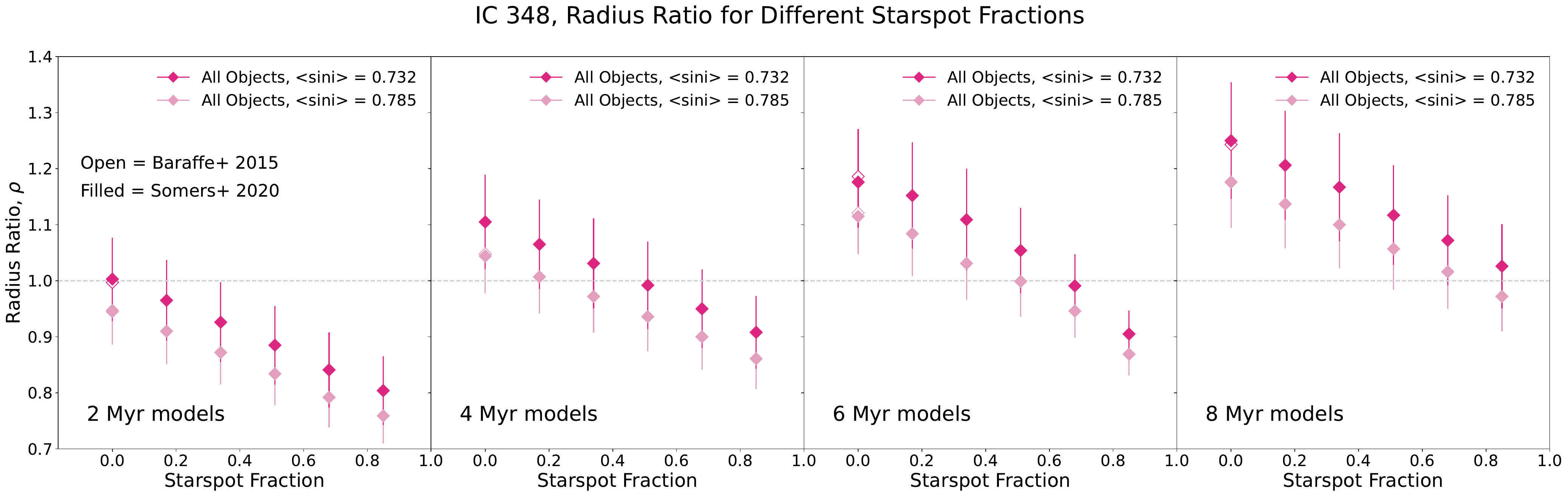}
    \caption{Radius ratio of radius from $v$~sin~$i$ and period measurements for stars in IC~348 to radius predictions from models with different starspot coverage fractions, assuming ages of 2, 4, 6, and 8 Myr.  Symbols are the same as in Figure \ref{fig:NAP_rsini_rho}.  The 2 Myr model is a good fit without starspots, which represents a lower limit on the age.  The 4 and 6 Myr models are best fit with a moderate amount of starspots, 51\% and 68\%, respectively. These are in good agreement with the updated age estimate for IC~348 of 5 $\pm$ 2 Myr from \cite{Luhman2024}, and coincide with results indicating that most TTSs may be at least 50\% covered in starspots \citep{Gangi2022, PerezPaolino2024}.} 
    \label{fig:IC348_rsini_rho}
\end{figure}

Fig. \ref{fig:IC348_rsini_rho} shows $\rho$ for our Model Comparison Sample stars in IC~348, with values reported in Tables \ref{tab:IC348_Baraffe_rho} and \ref{tab:IC348_SPOTS_rho}, located in the Appendix.  While the IC~348 ``all" group is small, the changes in $\rho$ are relatively consistent between starspot coverages and ages, as we have seen with larger samples, so we are confident in interpreting these results.  The 2 Myr starspot-free model has $\rho$ $\simeq$ 1 ($\rho$ = 0.997 $\pm$ 0.069); as we expect there to be some amount of starspots on these magnetically active stars, this places 2 Myr as a lower limit for the age.  4 Myr is an appropriate age for our IC~348 stars starting with a starspot coverage of only 17\%, up to slightly more than 68\%, and the same is true for 6 Myr starting at 34\%.  At 8 Myr, 85\% starspot coverage has $\rho$ in agreement with 1, but it is unlikely for IC~348 to be any older than this.  If we assume that most TTSs have an average starspot coverage of around 50\%$-$85\%, as reported by \cite{PerezPaolino2024}, then 4$-$6 Myr is the most appropriate age range for the cluster.  IC~348 was frequently estimated to be around 2$-$3 Myr (e.g., \citealt{Luhman1998, Luhman2003, LEL2016, Wang2022}), but our results are in closer agreement to recent revisions to an older age of 5 or 6 Myr \citep{Bell2013, Luhman2024}. 

\section{Conclusions} \label{sec:conclusion}

We have obtained high-resolution optical spectroscopy of a sample of low-mass PMS stars in IC~5070 and IC~348. 
We used cross-correlation techniques to measure the projected rotational velocity, $v$~sin~$i$, which allows us to constrain the minimum rotation velocity and, when combined with rotation periods, the minimum radius for these stars.  We supplemented these measurements with information from the literature about circumstellar disk status and binarity, which allowed us to explore the possible effects of these properties on the angular momentum by comparing the $v$~sin~$i$ distributions.  Due to the small size of our IC~5070 sample, we also created a combined sample with published results for stars in the ONC, a cluster with a similar age and thus a similar evolutionary stage.  We also used a maximum likelihood method to compare the average radius of stars in the clusters to radii for stars of similar luminosity predicted by stellar evolution models.  
We used the average radius ratio to compare models with different ages and starspot fractions, in order to constrain the ages of the clusters based on the stars in our sample.  The main findings of this work are summarized as follows:
\begin{enumerate}
    \item We compare the $v$~sin~$i$ distributions of the Class~II and Class~III stars in the combined IC~5070/ONC sample and find a statistically significant difference (K-S test $p$-value = 0.0043), with Class~III stars appearing to rotate faster.  In the IC~348 cluster, we see a difference in the distributions at the $\sim$10\% level (K-S test $p$-value = 0.091). While the Class~III stars above the velocity resolution limit were faster than the Class~II stars, we also noted that a higher proportion of Class~III stars were below our velocity resolution limit in IC~348 compared to both the IC~5070/ONC samples and the samples we analyzed in NGC 2264 \citep{Gray2024}. We theorize that a portion of these slow-rotating stars may have recently lost their disks but have not had enough time to spin up to higher rotation rates.
    \item Overestimation of $v$~sin~$i$ from cross-correlation is a higher risk in binary systems, where unresolved companions may broaden the CCF peak.  To minimize this, we created a sample that excluded objects with large variations in their $v$~sin~$i$ measurements that we attributed to binarity for use in our analysis of the $v$~sin~$i$ distributions.  We did not find a statistically significant difference between the $v$~sin~$i$ distributions of single stars and binary stars in either cluster, but the small size of our samples and our velocity resolution limit make it difficult to draw a firm conclusion.  We did observe that a higher fraction of the single stars than the binary stars were below our velocity resolution limit.  In our combined IC~5070/ONC sample, we found that there was a higher proportion of slow rotators for Class~II single stars compared to Class~II binary stars, similar to what we observed in our NGC 2264 sample \citep{Gray2024}.  This may support theories that a disruption to the disk or disk-locking processes due to binarity may be responsible for the faster rotation of binaries observed in other clusters.  However, the difference between fractions of slow rotators between Class~III single stars and Class~III binaries was only slightly smaller, indicating that influences on the disk may not be the only factor, although our smaller sample size may have also affected our results.
    \item We use a maximum likelihood method to compare geometrical radii (estimated from combining $v$~sini~$i$ and rotation period measurements) to radii predicted by stellar evolutionary models.  We explore a range of ages and starspot fractions for the sample of stars in each cluster to constrain lower and upper limits for the ages, and ultimately a best-fitting age estimate.  Based on evidence that TTSs are expected to be significantly ($>$50\%) covered in starspots, we estimate ages of 1$-$2 Myr for IC~5070 and 4$-$6 Myr for IC~348.  We also attempted to explore a possible mass-dependence on age estimates that has been observed in other clusters, where higher-mass stars appear to be older than lower-mass stars.  We split our combined IC~5070/ONC sample into three temperature subgroups as a proxy for mass divisions, but we did not see a mass-dependent age difference between our ``lower" and ``mid" groups, which consisted of stars that were below 4000 K, and mainly had masses $<$1.1 $M_{\odot}$.  The results for the ``upper" group were unstable, likely due to the small sample size, so we were unable to fully explore a possible mass-dependence in age estimates.
\end{enumerate}

\begin{acknowledgments}

We thank the anonymous reviewer for the thoughtful and detailed comments, which significantly improved the quality of the manuscript. L.M.G. acknowledges support from the Indiana University (IU) Astronomy Department Sullivan Fellowship and the IU College of Arts and Sciences Dissertation Research Fellowship.  We thank the Indiana University College of Arts and Sciences for funding IU's share of the WIYN telescope.  We also thank the staff of the WIYN Observatory and Kitt Peak National Observatory for their help and support during our WIYN Hydra observing runs.

This work has made use of data from the European Space Agency (ESA) mission
{\it Gaia} (\url{https://www.cosmos.esa.int/gaia}), processed by the {\it Gaia}
Data Processing and Analysis Consortium (DPAC,
\url{https://www.cosmos.esa.int/web/gaia/dpac/consortium}). Funding for the DPAC
has been provided by national institutions, in particular the institutions
participating in the {\it Gaia} Multilateral Agreement.  Some of the data presented in this paper were obtained from the Mikulski Archive for Space Telescopes (MAST).  This research has made use of NASA’s Astrophysics Data System (ADS) Abstract Service and of the SIMBAD database, operated at CDS, Strasbourg, France.  This research made use of the open-source Python package exoctk, the Exoplanet Characterization Toolkit \citep{Bourque2021}.

\end{acknowledgments}

\begin{contribution}


\end{contribution}

\facilities{WIYN(Hydra)}

\software{astropy \citep{Astropy2013,Astropy2018,Astropy2022}, IRAF \citep{Tody1986, Tody1993}}

\appendix


\begin{deluxetable}{cccc}[p]
\tablecaption{Radius ratio, $\rho$, between radii measured for low-mass PMS stars in IC~5070 and radii predicted by the \cite{Baraffe2015} evolutionary models at 1, 2, and 3 Myr. The analysis used 16 objects, 8 of which had a measured value of $r$~sin~$i$. We include $\rho$ calculated assuming an isotropic spin-axis distribution ($\langle$sin~$i$$\rangle$ = 0.785) and using the observed $\langle$sin~$i$$\rangle$ of the sample, 0.679.}
\label{tab:NAP_Baraffe_rho}
\tablehead{\colhead{} & \colhead{\textbf{Model Age}} & \colhead{$\rho$, $\langle$sin~$i$$\rangle$ = 0.785} & \colhead{$\rho$, $\langle$sin~$i$$\rangle$ = 0.679} }
\startdata
 & 1 Myr & 1.070 $\pm$ 0.034 & 1.155 $\pm$ 0.053 \\
 & 2 Myr & 1.186 $\pm$ 0.038 & 1.280 $\pm$ 0.059 \\
 & 3 Myr & 1.237 $\pm$ 0.043 & 1.340 $\pm$ 0.063 \\
\enddata
\end{deluxetable}

\begin{deluxetable}{ccccc}
\tablecaption{Radius ratio, $\rho$, between radii measured for low-mass PMS stars in IC~5070 and radii predicted by the \cite{Somers2020} evolutionary models at 1, 2, and 3 Myr. The analysis used 16 objects, 8 of which had a measured value of $r$~sin~$i$. We include $\rho$ calculated assuming an isotropic spin-axis distribution ($\langle$sin~$i$$\rangle$ = 0.785) and using the observed $\langle$sin~$i$$\rangle$ of the sample, 0.679.}
\label{tab:NAP_SPOTS_rho}
\tablehead{\multicolumn{2}{c}{\textbf{Model Age}} & \colhead{\textbf{Starspot Fraction}} & \colhead{$\rho$, $\langle$sin~$i$$\rangle$ = 0.785} & \colhead{$\rho$, $\langle$sin~$i$$\rangle$ = 0.679}  }
\startdata
\multicolumn{2}{c}{1 Myr} & 0\% & 1.071 $\pm$ 0.035 & 1.155 $\pm$ 0.053 \\
 & & 17\% & 1.031 $\pm$ 0.033 & 1.111 $\pm$ 0.051 \\
 & & 34\% & 0.992 $\pm$ 0.032 & 1.070 $\pm$ 0.049 \\
 & & 51\% & 0.952 $\pm$ 0.031 & 1.027 $\pm$ 0.047 \\
 & & 68\% & 0.911 $\pm$ 0.030 & 0.983 $\pm$ 0.045 \\
 & & 85\% & 0.864 $\pm$ 0.028 & 0.932 $\pm$ 0.041 \\\\
\multicolumn{2}{c}{2 Myr} & 0\% & 1.206 $\pm$ 0.039 & 1.302 $\pm$ 0.059 \\
 & & 17\% & 1.163 $\pm$ 0.038 & 1.255 $\pm$ 0.058 \\
 & & 34\% & 1.121 $\pm$ 0.036 & 1.209 $\pm$ 0.056 \\
 & & 51\% & 1.075 $\pm$ 0.035 & 1.161 $\pm$ 0.053 \\
 & & 68\% & 1.024 $\pm$ 0.033 & 1.106 $\pm$ 0.051 \\
 & & 85\% & 0.972 $\pm$ 0.032 & 1.050 $\pm$ 0.049 \\\\
\multicolumn{2}{c}{3 Myr} & 0\% & 1.303 $\pm$ 0.042 & 1.406 $\pm$ 0.064 \\
 & & 17\% & 1.261 $\pm$ 0.043 & 1.363 $\pm$ 0.061 \\
 & & 34\% & 1.209 $\pm$ 0.039 & 1.305 $\pm$ 0.060 \\
 & & 51\% & 1.167 $\pm$ 0.040 & 1.261 $\pm$ 0.056 \\
 & & 68\% & 1.079 $\pm$ 0.036 & 1.168 $\pm$ 0.055 \\
 & & 85\% & 1.051 $\pm$ 0.035 & 1.134 $\pm$ 0.052 \\
\enddata
\end{deluxetable}

\begin{deluxetable}{cccccc}
\tablecaption{Radius ratio, $\rho$, between radii measured for low-mass PMS stars in IC~5070/ONC and radii predicted by the \cite{Baraffe2015} evolutionary models at 1, 2, and 3 Myr. The numbers below each temperature group indicate the total number of objects in that group, with the number of objects with a measured value of $r$~sin~$i$ in parentheses.  We include $\rho$ calculated assuming an isotropic spin-axis distribution ($\langle$sin~$i$$\rangle$ = 0.785) and using the observed $\langle$sin~$i$$\rangle$ of the subgroup.}
\label{tab:NAP-ONC_Baraffe_rho}
\tablehead{\colhead{} & \colhead{} & \colhead{\textbf{All}} & \colhead{\textbf{Lower}} & \colhead{\textbf{Mid}} & \colhead{\textbf{Upper}} \\ 
\colhead{} & \colhead{ } & \colhead{72 (43)} & \colhead{29 (22)} & \colhead{31 (13)} & \colhead{12 (8)} \\
\colhead{} & \colhead{\textbf{Model Age}} & \colhead{$\rho$} & \colhead{$\rho$} & \colhead{$\rho$} & \colhead{$\rho$} }
\startdata
\multicolumn{2}{l}{\textit{Isotropic $\langle$sin~$i$$\rangle$ =}} & \textit{0.785} & \textit{0.785} & \textit{0.785} & \textit{0.785}\\
 & 1 Myr & 1.073 $\pm$ 0.026 & 0.867 $\pm$ 0.053 & 1.073 $\pm$ 0.033 & 1.138 $\pm$ 0.051 \\
 & 2 Myr & 1.189 $\pm$ 0.044 & 0.996 $\pm$ 0.059 & 1.190 $\pm$ 0.036 & 1.593 $\pm$ 0.069 \\
 & 3 Myr & 1.197 $\pm$ 0.036 & 1.059 $\pm$ 0.062 & 1.231 $\pm$ 0.038 & 1.238 $\pm$ 0.102 \\
\hline
\multicolumn{2}{l}{\textit{Observed $\langle$sin~$i$$\rangle$ =}} & \textit{0.694} & \textit{0.649} & \textit{0.726} & \textit{0.767} \\
 & 1 Myr & 1.141 $\pm$ 0.033 & 1.016 $\pm$ 0.065 & 1.119 $\pm$ 0.039 & 1.153 $\pm$ 0.056 \\
 & 2 Myr & 1.359 $\pm$ 0.038 & 1.168 $\pm$ 0.074 & 1.242 $\pm$ 0.043 & 1.610 $\pm$ 0.071 \\
 & 3 Myr & 1.272 $\pm$ 0.039 & 1.241 $\pm$ 0.079 & 1.284 $\pm$ 0.044 & 1.241 $\pm$ 0.106 \\
\enddata
\end{deluxetable}

\begin{deluxetable}{ccccccc}
\tablecaption{Radius ratio, $\rho$, between radii measured for low-mass PMS stars in IC~5070/ONC and radii predicted by the \cite{Somers2020} evolutionary models at 1, 2, and 3 Myr. The numbers below each temperature group indicate the total number of objects in that group, with the number of objects with a measured value of $r$~sin~$i$ in parentheses.  Some $\rho$ are left blank where ln $\widehat{\mathcal{L}}$ did not have a clear maximum or we could not estimate the uncertainty.  We include $\rho$ calculated assuming an isotropic spin-axis distribution ($\langle$sin~$i$$\rangle$ = 0.785) and using the observed $\langle$sin~$i$$\rangle$ of the subgroup.}
\label{tab:NAP-ONC_SPOTS_rho}
\tablehead{\colhead{} & \colhead{} & \colhead{} & \colhead{\textbf{All}} & \colhead{\textbf{Lower}} & \colhead{\textbf{Mid}} & \colhead{\textbf{Upper}} \\ 
\colhead{} & \colhead{} & \colhead{ } & \colhead{72 (43)} & \colhead{29 (22)} & \colhead{31 (13)} & \colhead{12 (8)} \\
\multicolumn{2}{c}{\textbf{Model Age}} & \colhead{\textbf{Starspot Fraction}} & \colhead{$\rho$} & \colhead{$\rho$} & \colhead{$\rho$} & \colhead{$\rho$} }
\startdata
\multicolumn{3}{l}{\textit{Isotropic $\langle$sin~$i$$\rangle$ = }} & \textit{0.785} & \textit{0.785} & \textit{0.785} & \textit{0.785} \\
\multicolumn{2}{c}{1 Myr} & 0\% & 1.076 $\pm$ 0.026 & 0.886 $\pm$ 0.052 & 1.070 $\pm$ 0.033 & 1.170 $\pm$ 0.047 \\
     & & 17\% & 1.029 $\pm$ 0.028 & 0.854 $\pm$ 0.051 & 1.030 $\pm$ 0.031 & 1.218 $\pm$ 0.076 \\
     & & 34\% & 0.993 $\pm$ 0.025 & 0.823 $\pm$ 0.049 & 0.992 $\pm$ 0.030 & 1.224 $\pm$ 0.142 \\
     & & 51\% & 0.941 $\pm$ 0.023 & 0.789 $\pm$ 0.047 & 0.952 $\pm$ 0.029 & 1.258 $\pm$ 0.095 \\
     & & 68\% & 0.875 $\pm$ 0.028 & 0.757 $\pm$ 0.045 & 0.911 $\pm$ 0.027 & 0.866 $\pm$ 0.068 \\
     & & 85\% & 0.850 $\pm$ 0.024 & 0.722 $\pm$ 0.043 & 0.867 $\pm$ 0.027 & 1.254 $\pm$ 0.081 \\\\
\multicolumn{2}{c}{2 Myr} & 0\% & 1.221 $\pm$ 0.029 & 0.957 $\pm$ 0.057 & 1.207 $\pm$ 0.036 & 1.324 $\pm$ 0.049 \\
     & & 17\% & 1.153 $\pm$ 0.031 & 0.921 $\pm$ 0.054 & 1.163 $\pm$ 0.035 & 1.681 $\pm$ 0.217 \\
     & & 34\% & 1.110 $\pm$ 0.026 & 0.882 $\pm$ 0.052 & 1.121 $\pm$ 0.034 & 1.205 $\pm$ 0.064 \\
     & & 51\% & 1.046 $\pm$ 0.014 & 0.845 $\pm$ 0.050 & 1.071 $\pm$ 0.033 & 1.625 $\pm$ 0.044 \\
     & & 68\% & 0.945 $\pm$ 0.053 & 0.806 $\pm$ 0.046 & 1.011 $\pm$ 0.030 & 0.995 $\pm$ 0.091 \\
     & & 85\% & 0.949 $\pm$ 0.028 & 0.769 $\pm$ 0.044 & 0.967 $\pm$ 0.031 & 1.439 $\pm$ 0.105 \\\\
\multicolumn{2}{c}{3 Myr} & 0\% & 1.317 $\pm$ 0.032 & 1.051 $\pm$ 0.063 & 1.302 $\pm$ 0.040 & 1.478 $\pm$ 0.077 \\
     & & 17\% & 1.214 $\pm$ 0.037 & 1.012 $\pm$ 0.061 & 1.256 $\pm$ 0.039 &              \\
     & & 34\% & 1.193 $\pm$ 0.028 & 0.968 $\pm$ 0.059 & 1.210 $\pm$ 0.036 & 1.259 $\pm$ 0.053 \\
     & & 51\% & 1.089 $\pm$ 0.051 & 0.927 $\pm$ 0.056 & 1.163 $\pm$ 0.036 &              \\
     & & 68\% & 1.029 $\pm$ 0.039 & 0.886 $\pm$ 0.054 & 1.070 $\pm$ 0.033 & 1.076 $\pm$ 0.089 \\
     & & 85\% & 1.062 $\pm$ 0.030 & 0.846 $\pm$ 0.051 & 1.054 $\pm$ 0.031 & 1.275 $\pm$ 0.051 \\
\hline
\multicolumn{3}{l}{\textit{Observed $\langle$sin~$i$$\rangle$ = }} & \textit{0.694} & \textit{0.649} & \textit{0.726} & \textit{0.767} \\
\multicolumn{2}{c}{1 Myr} & 0\% & 1.153 $\pm$ 0.035 & 1.039 $\pm$ 0.066 & 1.117 $\pm$ 0.038 & 1.181 $\pm$ 0.052 \\
     & & 17\% & 1.111 $\pm$ 0.037 & 1.002 $\pm$ 0.064 & 1.074 $\pm$ 0.036 & 1.233 $\pm$ 0.073 \\
     & & 34\% & 1.069 $\pm$ 0.034 & 0.965 $\pm$ 0.057 & 1.036 $\pm$ 0.035 & 1.262 $\pm$ 0.140 \\
     & & 51\% & 1.000 $\pm$ 0.032 & 0.927 $\pm$ 0.058 & 0.994 $\pm$ 0.033 & 1.286 $\pm$ 0.084 \\
     & & 68\% & 0.932 $\pm$ 0.027 & 0.888 $\pm$ 0.057 & 0.952 $\pm$ 0.031 & 0.876 $\pm$ 0.066 \\
     & & 85\% & 0.898 $\pm$ 0.023 & 0.847 $\pm$ 0.054 & 0.905 $\pm$ 0.032 & 1.262 $\pm$ 0.085 \\\\
\multicolumn{2}{c}{2 Myr} & 0\% & 1.310 $\pm$ 0.040 & 1.125 $\pm$ 0.070 & 1.259 $\pm$ 0.043 & 1.334 $\pm$ 0.053 \\
     & & 17\% & 1.229 $\pm$ 0.042 & 1.081 $\pm$ 0.069 & 1.213 $\pm$ 0.041 & 1.569 $\pm$ 0.040 \\
     & & 34\% & 1.185 $\pm$ 0.034 & 1.036 $\pm$ 0.065 & 1.170 $\pm$ 0.040 & 1.225 $\pm$ 0.073 \\
     & & 51\% & 1.109 $\pm$ 0.012 & 0.990 $\pm$ 0.060 & 1.117 $\pm$ 0.038 & 1.620 $\pm$ 0.041 \\
     & & 68\% & 1.031 $\pm$ 0.031 & 0.947 $\pm$ 0.057 & 1.054 $\pm$ 0.034 & 1.000 $\pm$ 0.093 \\
     & & 85\% & 1.007 $\pm$ 0.028 & 0.904 $\pm$ 0.057 & 1.008 $\pm$ 0.035 & 1.392 $\pm$ 0.080 \\\\
\multicolumn{2}{c}{3 Myr} & 0\% & 1.415 $\pm$ 0.047 & 1.234 $\pm$ 0.078 & 1.359 $\pm$ 0.046 & 1.505 $\pm$ 0.088 \\
     & & 17\% & 1.269 $\pm$ 0.006 & 1.188 $\pm$ 0.076 & 1.313 $\pm$ 0.044 &              \\
     & & 34\% & 1.269 $\pm$ 0.035 & 1.138 $\pm$ 0.072 & 1.263 $\pm$ 0.043 & 1.270 $\pm$ 0.057 \\
     & & 51\% & 1.202 $\pm$ 0.061 & 1.088 $\pm$ 0.069 & 1.215 $\pm$ 0.040 &              \\
     & & 68\% & 1.100 $\pm$ 0.034 & 1.040 $\pm$ 0.067 & 1.117 $\pm$ 0.037 & 1.081 $\pm$ 0.092 \\
     & & 85\% & 1.160 $\pm$ 0.044 & 0.990 $\pm$ 0.065 & 1.102 $\pm$ 0.037 & 1.289 $\pm$ 0.055 \\
\enddata
\end{deluxetable}

\begin{deluxetable}{cccc}
\tablecaption{Radius ratio, $\rho$, between radii measured for low-mass PMS stars in IC~348 and radii predicted by the \cite{Baraffe2015} evolutionary models at 2, 4, 6, and 8 Myr. The analysis used 18 objects, 7 of which had a measured value of $r$~sin~$i$. We include $\rho$ calculated assuming an isotropic spin-axis distribution ($\langle$sin~$i$$\rangle$ = 0.785) and using the observed $\langle$sin~$i$$\rangle$ of the sample, 0.732.}
\label{tab:IC348_Baraffe_rho}
\tablehead{ 
\colhead{} & \colhead{\textbf{Model Age}} & \colhead{$\rho$, $\langle$sin~$i$$\rangle$ = 0.785} & \colhead{$\rho$, $\langle$sin~$i$$\rangle$ = 0.732} }
\startdata
 & 2 Myr & 0.945 $\pm$ 0.056 & 0.997 $\pm$ 0.069 \\
 & 4 Myr & 1.048 $\pm$ 0.062 & 1.105 $\pm$ 0.075 \\
 & 6 Myr & 1.121 $\pm$ 0.069 & 1.186 $\pm$ 0.085 \\
 & 8 Myr & 1.176 $\pm$ 0.074 & 1.243 $\pm$ 0.092 \\
\enddata
\end{deluxetable}

\begin{deluxetable}{ccccc}
\tablecaption{Radius ratio, $\rho$, between radii measured for low-mass PMS stars in IC~348 and radii predicted by the \cite{Somers2020} evolutionary models at 2, 4, 6, and 8 Myr. The analysis used 18 objects, 7 of which had a measured value of $r$~sin~$i$. We include $\rho$ calculated assuming an isotropic spin-axis distribution ($\langle$sin~$i$$\rangle$ = 0.785) and using the observed $\langle$sin~$i$$\rangle$ of the sample, 0.732.}
\label{tab:IC348_SPOTS_rho}
\tablehead{
\multicolumn{2}{c}{\textbf{Model Age}} & \colhead{\textbf{Starspot Fraction}} & \colhead{$\rho$, $\langle$sin~$i$$\rangle$ = 0.785} & \colhead{$\rho$, $\langle$sin~$i$$\rangle$ = 0.732} }

\startdata
\multicolumn{2}{c}{2 Myr} & 0\%  & 0.947 $\pm$ 0.061 & 1.003 $\pm$ 0.074 \\
 &   & 17\% & 0.910 $\pm$ 0.059 & 0.965 $\pm$ 0.072 \\
 &   & 34\% & 0.872 $\pm$ 0.057 & 0.926 $\pm$ 0.071 \\
 &   & 51\% & 0.834 $\pm$ 0.056 & 0.885 $\pm$ 0.070 \\
 &   & 68\% & 0.792 $\pm$ 0.053 & 0.841 $\pm$ 0.067 \\
 &   & 85\% & 0.759 $\pm$ 0.049 & 0.804 $\pm$ 0.061 \\\\
\multicolumn{2}{c}{4 Myr} & 0\%  & 1.044 $\pm$ 0.066 & 1.105 $\pm$ 0.084 \\
 &   & 17\% & 1.007 $\pm$ 0.065 & 1.065 $\pm$ 0.080 \\
 &   & 34\% & 0.972 $\pm$ 0.064 & 1.031 $\pm$ 0.080 \\
 &   & 51\% & 0.936 $\pm$ 0.062 & 0.992 $\pm$ 0.078 \\
 &   & 68\% & 0.900 $\pm$ 0.058 & 0.950 $\pm$ 0.070 \\
 &   & 85\% & 0.861 $\pm$ 0.054 & 0.908 $\pm$ 0.065 \\\\
\multicolumn{2}{c}{6 Myr} & 0\%  & 1.115 $\pm$ 0.067 & 1.176 $\pm$ 0.081 \\
 &   & 17\% & 1.084 $\pm$ 0.076 & 1.152 $\pm$ 0.095 \\
 &   & 34\% & 1.031 $\pm$ 0.065 & 1.109 $\pm$ 0.091 \\
 &   & 51\% & 0.999 $\pm$ 0.063 & 1.054 $\pm$ 0.076 \\
 &   & 68\% & 0.946 $\pm$ 0.048 & 0.991 $\pm$ 0.056 \\
 &   & 85\% & 0.869 $\pm$ 0.038 & 0.905 $\pm$ 0.042 \\\\
\multicolumn{2}{c}{8 Myr} & 0\%  & 1.176 $\pm$ 0.082 & 1.250 $\pm$ 0.104 \\
 &   & 17\% & 1.137 $\pm$ 0.079 & 1.206 $\pm$ 0.097 \\
 &   & 34\% & 1.100 $\pm$ 0.078 & 1.167 $\pm$ 0.096 \\
 &   & 51\% & 1.057 $\pm$ 0.073 & 1.117 $\pm$ 0.089 \\
 &   & 68\% & 1.016 $\pm$ 0.066 & 1.072 $\pm$ 0.080 \\
 &   & 85\% & 0.972 $\pm$ 0.062 & 1.026 $\pm$ 0.075\\
\enddata
\end{deluxetable}

\bibliography{IC5070_IC348_bib}{}
\bibliographystyle{aasjournalv7}



\end{document}